# Optical bottle microresonators

M. Sumetsky

Aston Institute of Photonic Technologies, Aston University, Birmingham B4 7ET, UK

Email: m.sumetsky@aston.ac.uk

**Abstract**

The optical microresonators reviewed in this paper are called bottle microresonators because their profile often resembles an elongated spheroid or a microscopic bottle. These resonators are commonly fabricated from an optical fiber by variation of its radius. Generally, variation of the bottle microresonator (BMR) radius along the fiber axis can be quite complex presenting, e.g., a series of coupled BMRs positioned along the fiber. Similar to optical spherical and toroidal microresonators, BMRs support whispering gallery modes (WGMs) which are localized inside the resonator due to the effect of total internal reflection. The elongation of BMRs along the fiber axis enables their several important properties and applications not possible to realize with other optical microresonators. The paper starts with the review of the BMR theory, which includes their spectral properties, slow WGM propagation along BMRs, theory of Surface Nanoscale Axial Photonics (SNAP) BMRs, theory of resonant transmission of light through BMR microresonators coupled to transverse waveguides (microfibers), theory of nonstationary WGMs in BMRs, and theory of nonlinear BMRs. Next, the fabrication methods of BMRs including melting of optical fibers, fiber annealing in SNAP technology, rolling of semiconductor bilayers, solidifying of a UV-curable adhesive, and others are reviewed. Finally, the applications of BMRs which either have been demonstrated or feasible in the nearest future are considered. These applications include miniature BMR delay lines, BMR lasers, nonlinear BMRs, optomechanical BMRs, BMR for quantum processing, and BMR sensors.

## 1. Introduction

Over the last decades, there has been a significant interest to the investigation and applications of dielectric optical microresonators. Several books, book chapters and review papers summarize the achievements in this research area (see e.g., [1-10]). The goal of the present review is to outline the major theoretical and experimental results in the development of the bottle microresonators [11] which were investigated in several research groups worldwide over the last decade and became of special interest for applications in photonics and adjacent disciplines.

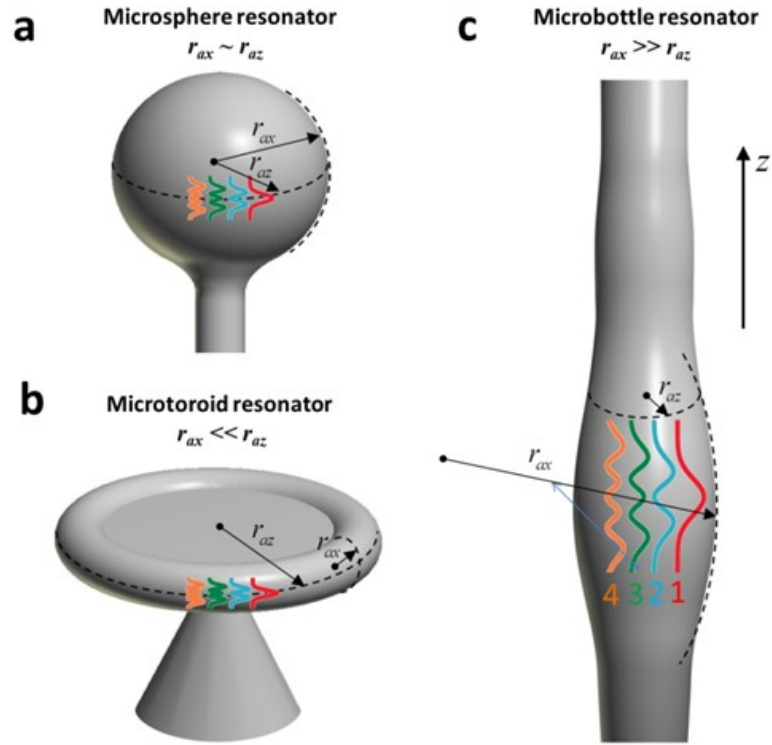

Fig. 1. Illustration of microsphere (a), microtoroid (b) and microbottle (c) resonators. (Reproduced with permission from Ref. [19]).

In order to understand the basic properties of a bottle microresonator (BMR) it is useful to compare them with the spherical and toroidal microresonators (Fig. 1). The major similarity of spherical, toroidal, and bottle microresonators is that optical states are localized inside these resonators due to the effect of total internal reflection and have the structure of a whispering gallery mode (WGM). These modes are propagating along the surface of these resonators and are analogous to acoustic WGMs discovered by Lord Rayleigh for acoustic waves a century ago [12] and proposed for fabrication of high Q-factor electromagnetic dielectric resonators by Richtmyer in 1939 [13]. The spherical microresonators since their experimental demonstration in paper [14] have a multidecade history of research and development [1-10]. The toroidal microresonators, which were introduced in papers [15, 16], have found applications in optical linear and nonlinear signal processing, lasing, sensing, and quantum networking. While the effect of localization of WGMs near the stable closed optical rays (geodesics) and, in particular, near the circumference of an elongated spheroidal microresonator, was known long ago [17], a BMR with arbitrary smooth radius variation along its axis was introduced and theoretically described in 2004 [11] following the experimental demonstration of such resonator in 2001 [18]. The major differences between spherical, toroidal, and BMRs are illustrated in Fig. 1 [19]. Assuming for simplicity that these resonators are axially symmetric, we can characterize their characteristic dimensions by axial and azimuthal radii, $r_{ax}$ and $r_{az}$. For spherical microresonators (Fig. 1(a)), $r_{ax}$ and $r_{az}$ are comparable, $r_{ax} \sim r_{az}$. For toroidal microresonators (Fig. 1(b)), the axial radius is relatively small, $r_{ax} << r_{az}$. In contrast, for BMRs (Fig. 1(c)), the axial radius can be made much greater than the azimuthal radius, $r_{ax} >> r_{az}$. For example, BMRs fabricated in SNAP (Surface Nanoscale Axial Photonics) technology [19-28] may have the axial radius exceeding a kilometre [23, 24]. For this reason, the characteristic variation length of WGMs along the axis of a BMR can be much greater than that of spherical and toroidal microresonators. Besides interesting properties of slow WGMs described below, this fact significantly simplifies the access to these WGMs from the outside and makes the BMR very attractive for applications.

The paper starts with the review of the BMR theory (Section 2), which includes BMR spectral properties [11, 29, 30], slow WGM propagation along BMRs, theory of Surface Nanoscale Axial Photonics (SNAP) BMRs [20, 21, 25], theory of resonant transmission of light through BMR microresonators coupled to transverse waveguides (microfibers) [25], theory of nonstationary WGMs in BMRs [31], and theory of nonlinear BMRs [32-36]. Next, in Section 3, the fabrication methods of BMRs including melting of optical fibers, fiber annealing in SNAP technology, rolling of semiconductor bilayers, solidifying of a UV-curable adhesive, local heating, femtosecond laser inscription, and others are reviewed [18, 23, 37-57]. Finally, in Section 4, applications of BMRs which either have been demonstrated or seem feasible in the nearest future are considered. These applications include miniature BMR delay lines and dispersion compensators [24, 58-60], BMR lasers [61-68], nonlinear BMRs [69-75], optomechanical BMRs [76-83], BMR for quantum processing [84-88], and BMR sensors [89-97].

## 2. Theory of BMR

### 2.1. Semiclassical theory of BMRs

In the original paper [11] the behavior of WGMs in BMRs was considered in the semiclassical approximation. This approximation is valid for all modes except for those localized near the circumference with maximum radius of the resonator, in particular, for resonators with very small radius variation considered in SNAP. In this Section we will describe the WGM semiclassically and postpone the more accurate consideration to Section 2.3.

The characteristic structure of classical rays in a BMR is illustrated in Fig. 2. In cylindrical coordinates $(z, \rho, \varphi)$ these rays can be separated into the rays moving in the plane $(\rho, \varphi)$ normal to the fiber axis $z$ (Fig. 2(a)) and the rays moving in the plane $(z, \rho)$ (Fig. 2(b)). Due to the similarity of the shape of WGMs built on these rays to the magnetic bottles in plasma fusion [98], the optical microresonator shown in Fig. 2 was called a whispering gallery bottle in Ref. [11]. Often, this type of optical resonator is simply called the BMR.

In the semiclassical approximation, the WGMs in a BMR are described under the assumption that this resonator is axially symmetric and its wall determined by the equation $\rho = \rho_w(z)$ is weakly curved. Then, the wavelength eigenvalues $\lambda = \lambda_{mpq}$ of this resonator are determined by the quantization rule:

$$\int_{z_1}^{z_2} \beta_{mp}(z, \lambda) dz = \pi (q + \tfrac{1}{2}), \quad \beta_{mp}(z, \lambda) = \sqrt{\frac{4\pi^2 n_r^2}{\lambda^2} - \frac{\mu_{mp}^2}{\rho_w^2(z)}} \tag{1}$$

Here $n_r$ is the refractive index of the resonator (optical fiber) material, $z_1$ and $z_2$ are zeros of the square root under the integral (turning points) and constant $\mu_{mp}$ is the parameter which can approximately defined through the azimuthal and radial quantum numbers, $m$ and $p$, as a root of the Bessel function $J_m(\mu_{mp}) = 0$. For modes localized near the surface of the fiber, $p \ll m$ and $\mu_{mp} \approx m$. BMRs can be extremely elongated so that the eigenvalue separation between the azimuthal and radial quantum numbers $m$ and $p$ is sparse compared to the separation between the axial quantum numbers $q$.

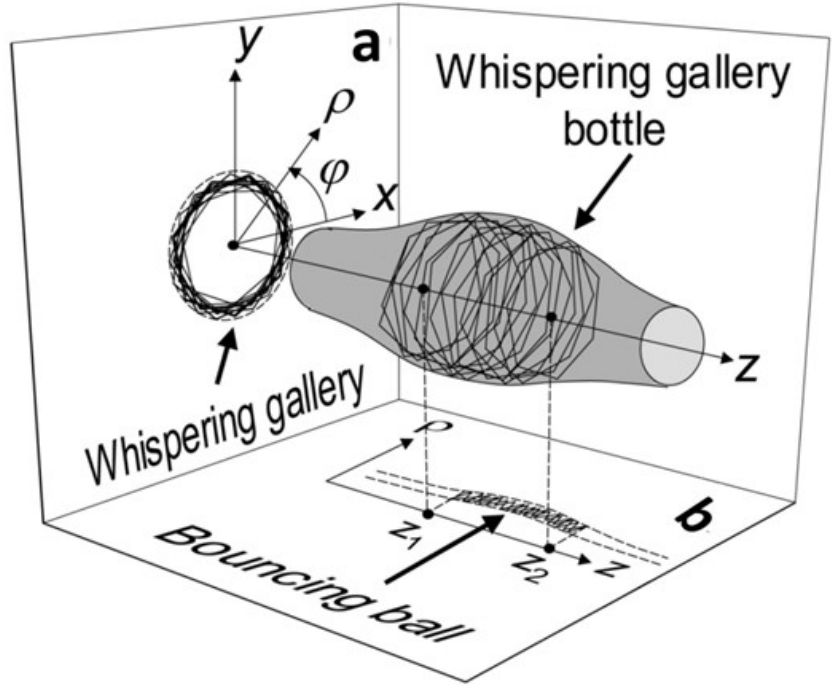

Fig. 2. Illustration of a BMR. (a) – projection of the localized WGM on the plane $(\rho, \varphi)$ and (b) – projection of the WGM on the plane $(z, \rho)$. (Reproduced with permission from Ref. [11]).

It has been shown in [11] that the shape of the BMR $\rho_w(z)$ can be determined from the given dependence of propagation constant $k_{mpq} = 2\pi n_r / \lambda_{mpq}$ on $q$ by the exact solution of Eq. (1). In particular, it was shown that there exist a resonator which propagation constant (frequency) eigenvalues are *equally spaced* along $q$. The shape $\rho_w(z)$ corresponding to equally spaced propagation constant eigenvalues with separation $dk_{mpq}/dq \equiv \Delta k$ has been found in [11] in the following simple form:

$$\rho_w(z) = \rho_0 \, |\cos(\Delta k z)| \qquad (2)$$

The interesting property of this shape is that it is independent of $m$ and $p$. Substitution of Eq. (2) into Eq. (1) yields the quantization rule:

$$k_{mpq} = \frac{2\pi n_r}{\lambda_{mpq}} = \frac{\mu_{mp}}{\rho_0} + (q + \tfrac{1}{2})\Delta k \ . \tag{3}$$

Remarkably, Eq. (3) shows that a BMR with the radius variation determined by Eq. (2) can serve as a 3D etalon, similar to the Fabry-Perot etalon in one dimension and used for several important applications such as miniature delay lines and frequency comb generators.

Generally, the variables in the wave equation for the BMR cannot be separated. For this reason, the behavior of the WGBs ray dynamics in this resonator may be stochastic [99]. In [11], in order to understand the ray behavior in a BMR, the Poincare surfaces of sections [99] for the non-prolate ($\Delta k\rho_0 = 1$) and prolate ($\Delta k\rho_0 = 0.1$) cosine-shaped microcavities were investigated for different values of dimensionless angular momentum $M = L_z/(v_0\rho_0) < 1$ expressed through the velocity $v_0$, and angular momentum $L_z$ of a particle propagating along the ray. This angular momentum conserves due to the axial symmetry. It was shown that, for the non-prolate cavity, the motion can be stochastic for small values of $M$ and becomes regular for larger $M$ when the rays localize near the plane of symmetry $z = 0$. However, for the strongly prolate cavity the stochasticity was shown to be strongly suppressed.

For shallow BMRs, the semiclassical description can be extended to small axial quantum numbers $q$. Then, the WGM propagation along the BMR axis is described by the one-dimensional wave equation analogous to the Schrödinger equation [20-25]. In particular, if the BMR profile can be approximated near the maximum radius $\rho_0$ by the quadratic dependence, $\rho_w(z) = \rho_0(1 - \Delta k^2 z^2 / 2)$, the axial dependence of WGMs is determined as $H_q((2\pi n_r\Delta k / \lambda)^{1/2}z)$ where $H_q(x)$ is the Hermite polynomial, $\lambda$ is the radiation wavelength, and $n_r$ is the refractive index of the fiber [29]. From Eq. (2), this approximation determines the equally spaced frequency eigenvalues of the BMR if $\Delta k^2 z^2 << 1$. Within this approximation, the number $q$ of equally spaced frequency eigenvalues can be still very large for very elongated resonators (see, e.g., [24, 33, 34]).

## 2.2. Phase velocity, group velocity, and slow light in BMRs

The characteristic feature of BMRs is their elongation along axis $z$ (Fig. 1). In the adiabatic approximation, when variation of the BMR radius $\rho_w(z)$ is sufficiently smooth and slow, the expression for the WGM can be written as [11]

$$U(\rho, z, \varphi) \approx \beta_{mp}(z, \lambda)^{-1/2} \exp\left(\pm i\int^z \beta_{mp}(z, \lambda)dz \pm im\phi\right)\Phi_{mp}\left(\frac{\rho}{\rho_w(z)}\right) \tag{4}$$

where the propagation constant $\beta_{mp}(z, \lambda)$ is determined by Eq. (1). Near the turning points $z_1$ and $z_2$ where $\beta_{mp}(z, \lambda) \to 0$ the dependence on $z$ in Eq. (4) should be replaced by more accurate expression through Airy functions [17]. In these regions, the propagation of light along axis $z$ is slow. In fact, the phase velocity

$$v_{ph}(z, \lambda) = \frac{2\pi c}{\lambda \beta_{mp}(z, \lambda)} \tag{5}$$

in these regions tends to infinity, while the group velocity

$$v_{gr}(z,\lambda) = -\frac{2\pi c}{\lambda^2}\left(\frac{\partial \beta_{mp}(z,\lambda)}{\partial \lambda}\right)^{-1} \qquad (6)$$

tends to zero. Calculation of the phase and group velocities using Eqs. (1), (5) and (6) yields the remarkable relation between these velocities and the speed of light $c$:

$$v_{gr}(z,\lambda)v_{ph}(z,\lambda) = \left(\frac{c}{n_r}\right)^2 \qquad (7)$$

From this equation, small group velocities (compared to the speed of light in the resonator material) correspond to large phase velocities.

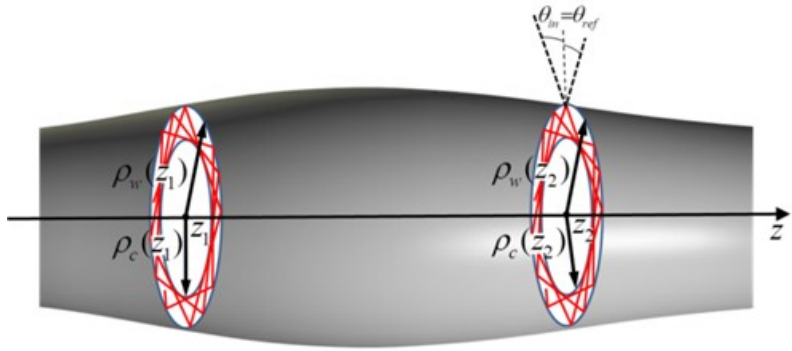

Fig. 3. Illustration of the behavior of classical rays near the turning points of the BMR.

The principal difference between the propagation of WGMs of our interest along the axis of an optical fiber (and, in particular, of a BMR) and the propagation of modes with relatively small azimuthal quantum numbers $m$ in an optical fiber (i.e., conventional propagation of light in a single mode fiber) should be emphasized. For relatively small $m$, slow propagation of light in optical fibers is impossible due to the cutoff effect which requires that the propagation constant $\beta$ is larger or around $2\pi/\lambda$ [100]. Alternatively, if $m$ is large enough, the zero value of propagation constant can be achieved. Let us derive the condition of $\beta = 0$ for a WGM in the semiclassical approximation, i.e., for large azimuthal and radial quantum numbers $m, p \gg 1$. To this end note that near the turning points $z_i$ the classical rays in a BMR are located close to its transverse cross-sections between the BMR wall of radius $\rho_w(z_i)$ and caustic of radius $\rho_c(z_i)$ (Fig. 3). To avoid the radiation of corresponding WGMs through the BMR surface, the angle of incidence of these rays $\theta_{in} = \arcsin(\rho_c/\rho_w)$, equal to the reflection angle $\theta_{ref}$, should be larger than the critical angle $\theta_c = \arcsin(n_r^{-1})$ for the total internal reflection (here we assume that the resonator is positioned in air with refractive index 1). On the other hand the semiclassical quantization rules in the cross-sections at $z = z_i$ can be written as [17]

$$\frac{2\pi n_r}{\lambda}\rho_c = m,$$

$$\frac{2\pi n_r}{\lambda}\left(\sqrt{\rho_w^2 - \rho_c^2} - \rho_c \arccos\left(\frac{\rho_c}{\rho_w}\right)\right) = \pi(p+\gamma),$$

(8)

where $\gamma \sim 1$ and $p, m \gg 1$. The WGM can reach the zero propagation constant before the cutoff if its radial quantum number is smaller than that corresponding to the rays with the critical angle of incidence. At the critical angle, we have $\rho_w = n_r \rho_c$. Then, from Eqs. (8) the WGMs with radial quantum numbers

$$p < \frac{m}{\pi}\left(\sqrt{n_r^2 - 1} - \arccos\frac{1}{n_r}\right)$$

(9)

can have the zero propagation constant. For silica BMRs with $n_r = 1.46$, this equation yields $p < 0.08m$. Assuming $\rho_w \sim 20\ \mu m$ [23] we have $m \sim 100$ and, thus, $p < 8$.

### 2.3.    Theory of weakly nonuniform BMRs

The WGM structure of a BMR can be very complex. However, this structure is simplified and can be described analytically in the case of a shallow resonator having relatively small radius variation $\rho_w(z) - \rho_0$. The most convenient way to fabricate such resonators is to slightly deform an originally uniform optical fiber. However, spatial deformation is not the only possible way to achieve the full localization of WGMs in an optical fiber. Therefore here we consider WGMs in an optical fiber possessing both radius and refractive index variations. Due to the small value of these variations, the expression for WGMs can be found in the cylindrical coordinates $(\rho, z, \varphi)$ in the separable form,

$$U(\rho, z, \varphi) = \exp(im\varphi)Q_{mp}(\rho)\Psi_{mp}(z).$$

(10)

Here function $Q_{mp}(\rho)$ satisfies the differential equation of the conventional fiber theory with zero propagation constant [3],

$$\frac{d^2 Q_{mp}(\rho, \lambda)}{d\rho^2} + \frac{1}{\rho}\frac{dQ_{mp}(\rho)}{d\rho} + \left[\left(\frac{2\pi n(\rho)}{\lambda_{mp}(z)}\right)^2 - \frac{m^2}{\rho^2}\right]Q_{mp}(\rho) = 0,$$

(11)

and $\Psi_{mp}(z)$ satisfies the one-dimensional wave equation [20-25]

$$\frac{d^2\Psi_{mp}}{dz^2} + \beta_{mp}^2(z)\Psi_{mp} = 0$$

(12)

with spatially dependent propagation constant

$$\beta_{mp}(z) = \frac{2^{3/2}\pi n_r}{\lambda_{mp}^{3/2}}\left(\lambda_{mp}(z) - \lambda\right)^{1/2}.$$

(13)

In Eqs. (11)-(13), function $\lambda_{mp}(z)$ is the cutoff wavelength of the optical fiber which corresponds to the local zero value of propagation constant $\beta_{mp}(z) = 0$. Crucially, all weak radial and refractive index dependencies on the axial coordinate $z$ are accumulated in the cutoff wavelength $\lambda_{mp}(z)$ which is expressed through the cutoff frequency $\omega_{mp}(z)$ as $\lambda_{mp}(z) = 2\pi c / \omega_{mp}(z)$. In Eq. (11), the dependence $\lambda_{mp}(z)$ is assumed to be adiabatic and parametric. Alternatively, in Eqs. (12) and (13) this dependence nontrivially determines the behavior of the WGMs along the resonator axis $z$. As an example of application of Eq. (12), we refer the reader to Fig. 10(d) to be considered in Section 3.7 of this review. This figure shows the spectrogram of 30 coupled BMRs calculated with Eq. (12) in excellent agreement with the experiment.

The phase velocity $v_{ph}(z)$ and group velocity $v_{gr}(z)$ of the WGM propagation along the BMR axis can be calculated from Eq. (7):

$$v_{ph}(z) = \frac{2\pi c}{\lambda_{mp}\beta_{mp}(z)} = \frac{c}{2^{1/2}n_r}\left(\frac{\lambda_{mp}}{\lambda_{mp}(z) - \lambda}\right)^{1/2} \tag{14}$$

$$v_{gr}(z) = -\frac{2\pi c}{\lambda_{mp}^2}\left(\frac{\partial\beta_{mp}(z)}{\partial\lambda}\right)^{-1} = 2^{1/2}\frac{c}{n_r}\left(\frac{\lambda_{mp}(z) - \lambda}{\lambda_{mp}}\right)^{1/2} = \frac{c^2}{n_r^2 v_{ph}(z)}. \tag{15}$$

### 2.4. Transmission amplitude through BMRs

Usually, light is launched in and collected from a BMR using a biconical optical fiber taper with a micron diameter waist (microfiber). Generally, there may be several input-output tapers oriented transversely to the BMR under test as illustrated in Fig. 4.

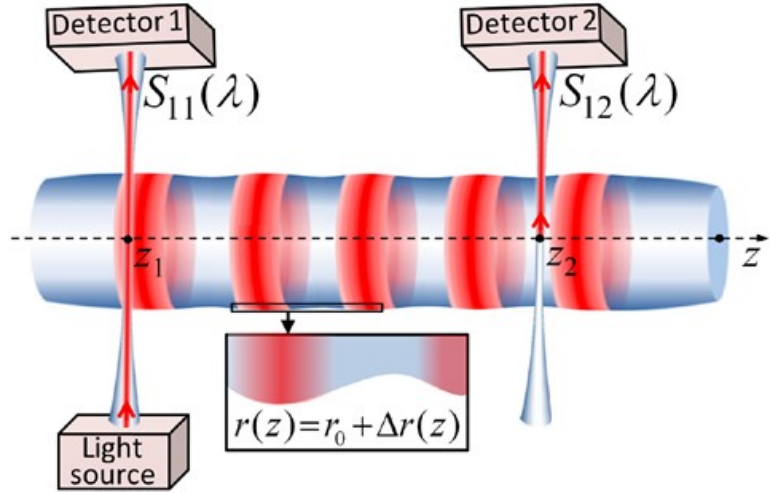

Fig. 4. A shallow BMR coupled to two biconical tapers with a micron diameter waist (Reproduced with permission from Ref. [25]).

The theory of resonant propagation of light through a shallow BMR, or a SNAP microresonator, coupled to input-output microfibers was developed in Ref. [25]. In the presence of $N$ microfibers coupled to the BMR at positions $z_n$, $n = 1, 2, ..., N$, along the BMR axis $z$, Eq. (12), which describes the dependence of WGMs on $z$, is modified to

$$\frac{d^2\Psi_{mp}}{dz^2} + \left(\beta_{mp}^2(z) + \sum_{n=1}^{N} D_n \delta(z - z_n)\right)\Psi_{mp} = 0 \qquad (16)$$

Here, complex parameter $D_n$ determine the effect of coupling to microfiber $n$ and $\delta(x)$ is the delta-function. The real part of $D_n$ determines the phase shift and its imaginary part determines losses due to coupling to the microfiber $n$. Modelling of coupling to the waveguides by zero-range potentials $D_n \delta(z - z_n)$ in Eq. (16) is justified for slow WGMs. In fact, while the microfiber diameter is ~1 μm, the characteristic axial wavelength of such WGMs usually exceeds 10 μm [25].

Assume that the microfiber with $n = 1$ serves as the input and output waveguide, while all other microfibers are the output waveguides only. Then the transmission amplitude from the input microfiber 1 to the output microfiber $n$ is

$$S_{1n} = S_{1n}^{(0)} - iC_1 C_n^* \overline{G}(z_1, z_n) . \qquad (17)$$

Here $\overline{G}(z_1, z_n)$ is the Green's function of Eq. (16) and $C_n$ are coupling parameters which together with $D_n$ fully determine the coupling between the microfiber and BMR. The Green's function $\overline{G}(z_1, z_n)$ can be expressed through the Green's function $G(z_1, z_n)$ of Eq. (16) in the absence of attached microfibers, i.e., for $D_n = 0$. The expressions of $\overline{G}(z_1, z_n)$ through $G(z_1, z_n)$ for $N = 1$ and $N = 2$ is given in [25]. Parameters $S_{1n}^{(0)}$ in Eq. (17) as well as $C_n$ and $D_n$ are weak function of radiation wavelength and usually can be considered as constants in the transmission bandwidth under interest. In the absence of losses we have $S_{1n}^{(0)} = 0$ for $n \geq 1$, $S_{11}^{(0)} = 1$, and $\text{Im}(D_n) = |C_n|^2/2$. Several particular cases when the Green's function of Eq. (16) and transmission amplitudes between tapers coupled to the BMR can be found analytically and numerically were considered in Ref. [25]. In the case of a single input-output microfiber, Eq. (17) was applied to the calculation of transmission amplitudes through a single BMR and coupled BMRs in [23-28, 51, 52] in excellent agreement with the experiment.

### 2.5. Nonstationary WGMs in BMRs and miniature optical buffer

There is a straightforward correspondence between the one-dimensional Schrödinger equation and the wave equation Eq. (12) describing propagation of light in a SNAP BMR. This correspondence becomes more transparent for non-stationary problems when the WGM can be presented in the form $\exp(im\varphi)Q_{mp}(\rho)\Psi_{mp}(z,t)$ generalizing Eq. (10). The equation that determines this propagation has the form of the non-stationary one-dimensional Schrödinger equation [31]

$$i\mu\frac{\partial\Psi}{\partial t} = -\frac{\partial^2\Psi}{\partial z^2} + V(z,t)\Psi . \qquad (18)$$

Here $\mu = 2\omega_0 n_0^2 / c^2$ and potential $V(z,t) = -8\pi^2 n_r^2 \lambda_{mp}(z,t) / \lambda_{mp}^3(z_0,t_0)$. It is assumed that variation $\lambda_{mp}(z,t) - \lambda_{mp}(z_0,t_0)$ is small (typically, it corresponds to the nanoscale variation of the optical fiber radius). The potential in Eq. (16) can be shifted by an arbitrary constant and in particular it can be set equal to $\beta_{mp}^2(z,t)$ calculated from Eq. (13) with time-dependent cutoff wavelength $\lambda_{mp}(z,t)$.

Eq. (16) allows to model a SNAP BMR proposed in Ref. [31] where it was shown that a tunable harmonic resonator illustrated in Fig. 5, can trap an optical pulse completely, hold it as long as the material losses permit, and release without distortion. Light is coupled in and out of this resonator through a transverse microfiber taper as described in Section 2.4 (Fig. 5(a)). The buffering process consists of three steps. First, the resonator is opened by nanoscale variation of its effective radius or refractive index (using, e.g., the driving pulse of the applied laser or electrical field) to let the optical pulse in (Fig. 5(b)). Next, the driving pulse is turned off when the optical pulse is completely inside the resonator which holds it for the duration of the required time delay (Fig. 5(c)). Finally, the optical pulse is released by opening the BMR with the driving pulse similar to Fig. 5 (b) (Fig. 5(d)). Fig. 5(e) shows the distribution of the field of the optical pulse, which is captured, held, and released by the described buffer. It is seen that the output pulse (top) exhibits the negligible change compared to the input pulse (bottom).

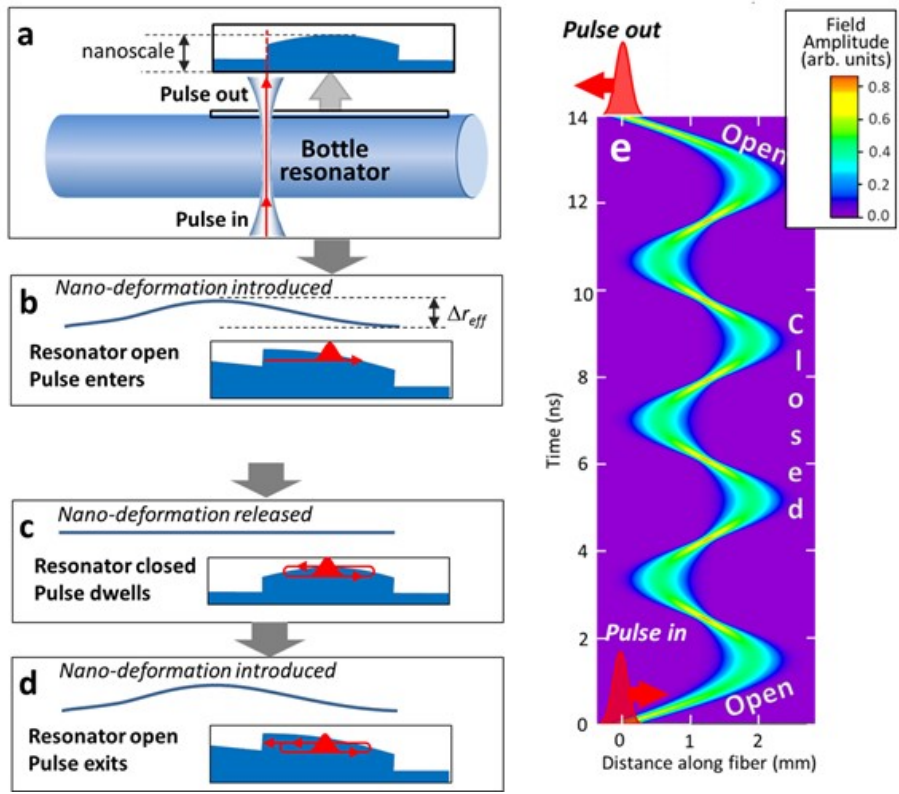

Fig. 5. Illustration and modelling of a SNAP BMR optical buffer. (a) – A SNAP BMR. The resonator is coupled to the transverse optical fiber taper. (b) – The switching nano-deformation introduced by the driving pulse of the applied laser or electrical field transfers the closed parabolic resonator into the open semi-parabolic resonator. (c) – After the driving pulse is turned off, the BMR

restores its original parabolic shape. The optical pulse oscillates inside it. (d) – Finally, the nano-deformation is introduced again and the pulse is released into the input-output waveguide. (e) – Distribution of the field of the optical pulse, which is captured, held, and released by the buffer, as a function of the coordinate along the BMR and time. (Reproduced with permission from Ref. [31]).

### 2.6. Nonlinear BMRs and frequency comp generation

The nonlinear Schrödinger equation describing propagation of WGMs in a BMR was independently introduced in Refs. [32, 34] and [35]. Here we will follow Refs. [32, 34] where the nonlinear Schrödinger equation for WGMs with fixed azimuthal quantum number $m$ was considered. This equation is generated from Eq. (18) by adding the term $\alpha |\Psi|^2$ describing the nonlinear Kerr effect to the potential:

$$i\mu \frac{\partial \Psi}{\partial t} = -\frac{\partial^2 \Psi}{\partial z^2} + \left( V(z,t) + \alpha |\Psi|^2 \right) \Psi \qquad (19)$$

Notice that in the presence of the input-output microfibers this equation should contain additional short-range complex potential terms similar to those in Eq. (16).

The transverse input-output microfiber was assumed to couple light in and out of BMR as illustrated Fig. 6(a). The potential considered in Refs. [32, 34] corresponded to the parabolic effective radius variation of the BMR (Fig. 6(b)) with maximum effective radius variation of 2.8 nm and gigantic axial radius of 1.6 km, which coincide with the SNAP BMR parameters experimentally demonstrated in Ref. [23]. It was shown in Refs. [32, 34, 35] that for relatively strong input field, the nonlinear effects described by Eq. (19) will lead to the generation of an optical frequency comb with an ultra-fine spectral spacing. The regimes of stable or quasiperiodic comb dynamics due to soliton excitation were identified. It was also shown that engineering of the BMR radius profile can be used to compensate for nonlinearity-induced dispersion.

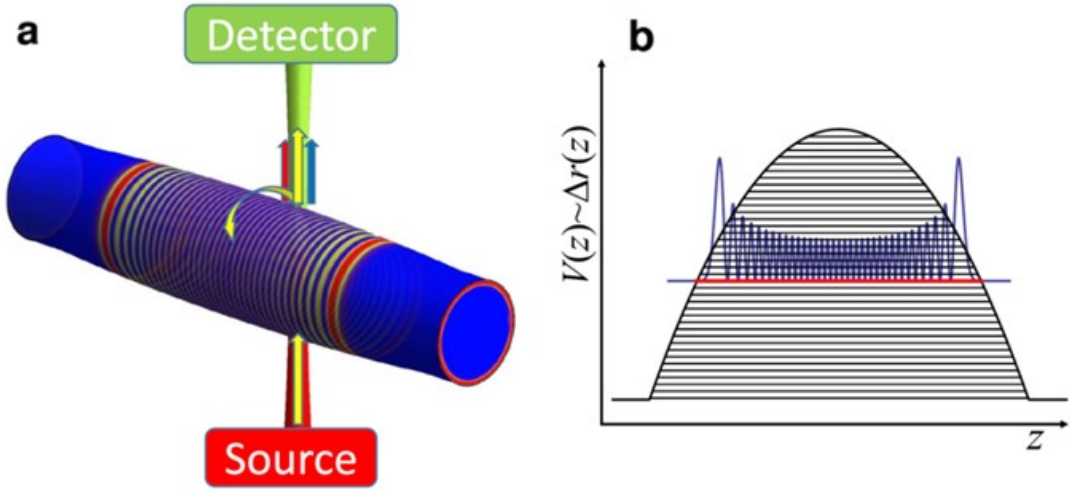

Fig. 6. (a) – Illustration of a SNAP BMR coupled to a microfiber. Characteristic intensity distributions of a WGM at the resonator surface and cross section are shown with color shading, where blue corresponds to zero, and red

corresponds to the maximum value. (b) – Effective potential corresponding to the parabolic radius variation. Blue line: the axial distribution of a WGM eigenstate [34]. (Reproduced with permission from Ref. [34]).

The numerical simulations based on Eq. (19) used the experimentally relevant physical parameters of BMR radius of 19 μm and radiation wavelength of $\lambda = 1.5$ μm, and silica refractive index $n_r = 1.46$. As an example, Fig. 7 shows the time evolution of the transmission spectrum in the logarithmic scale with a cutoff $10^{-10}$ [34]. It is seen that the comb is formed by approximately fifteen modes with sub-gigahertz frequency spacing, which remain stable over time. Since the SNAP BMR potential can be designed and introduced using the SNAP BMR fabrication methods reviewed below, compensation of the nonlinear dispersion and control the comb dynamics is possible.

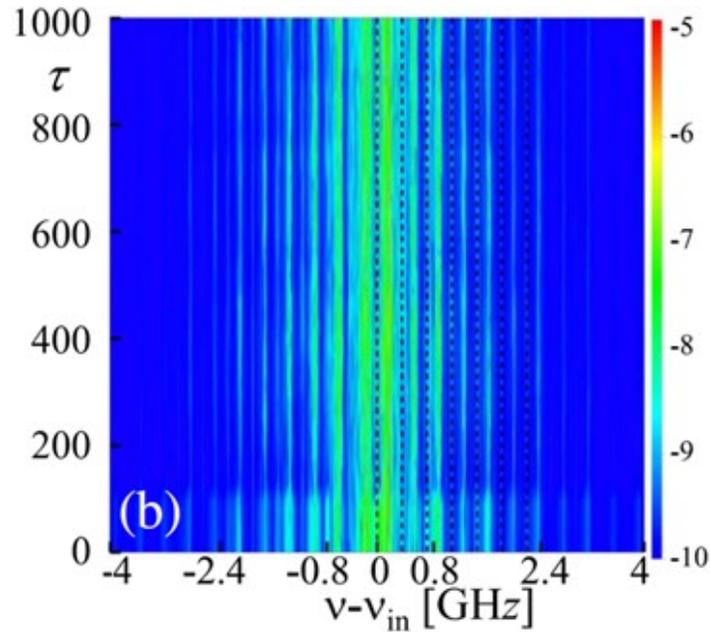

Fig. 7. Generation of frequency combs in the SNAP bottle resonator for the input frequency equal to the resonant frequency of the BMR state shown in Fig. 6(b). Black dashed lines indicate the eigenfrequencies of the parabolic potential [34]. (Reproduced with permission from Ref [34]).

## 3. Fabrication and characterization of BMRs

### 3.1. Melting and splicing optical fibers

Conventionally, BMRs are fabricated of optical fibers. The simplest approach consists in thinning the fiber at two positions so that the thicker region in between these positions forms a BMR. Usually, thinning is performed by melting the fiber in a fusion splicer, flame, or using a $CO_2$ laser. In the first demonstration of BMRs [18] and in Refs. [37, 40] they were fabricated by pulling the optical fiber melted by the $CO_2$ laser beam or flame (Fig. 8(a)). In Ref [39], BMRs were fabricated using the fiber splicing method (Fig. 8(b)). Fiber splicing is the process in which the ends of two optical fibers, which are cleaved and aligned, are heated to a melting temperature and fused together. The authors of Ref.

[39] used the fiber splicer to develop the "soften-and-compress" technique for the BMR fabrication. They softened a piece of continuous fiber positioned in the fiber splicer while simultaneously compressing it. As the result, they fabricated a BMR which shape was determined by the melting temperature and the applied compression. A modified approach of using a fiber splicer to create BMRs was explored in Ref. [48].

### 3.2. Rolling of semiconductor bilayers

A completely different BMR fabrication method was developed in Refs. [38, 45]. The authors of Ref. [38] demonstrated a BMR which was created by rolling of semiconductor bilayers which have nanometer scale thickness and lifted-off from the substrate (Fig. 8(c)). The layer thickness was 50 nm only, while the BMR was formed of less than a three layer roll. The properties of the created BMR are similar to regular BMRs. Since this microresonator was formed of an empty tube, the authors suggested that the term "empty-bottle resonator" would be more precise in this case. The most important feature of this microresonator is the structured rolling edge with a parabolic lobe which turns the structure into a BMR with parabolic effective radius variation. This structure introduces the parabolic variation of the effective tube radius, which, in analogy to the theory of SNAP resonators considered in Section 2.3, results in formation of a BMR.

### 3.3. Hollow BMR (bubble microresonators)

In Refs. [41, 42], fabrication of a hollow BMR (bubble microresonators) with micron wall thickness was demonstrated. The blowing of this microresonator was performed by pressurizing it internally and softening with a $CO_2$ laser (Fig. 8(d)). Alternatively, the authors of Ref. [43] used a fiber splicer to fabricate a hollow BMR from silica microcapillary.

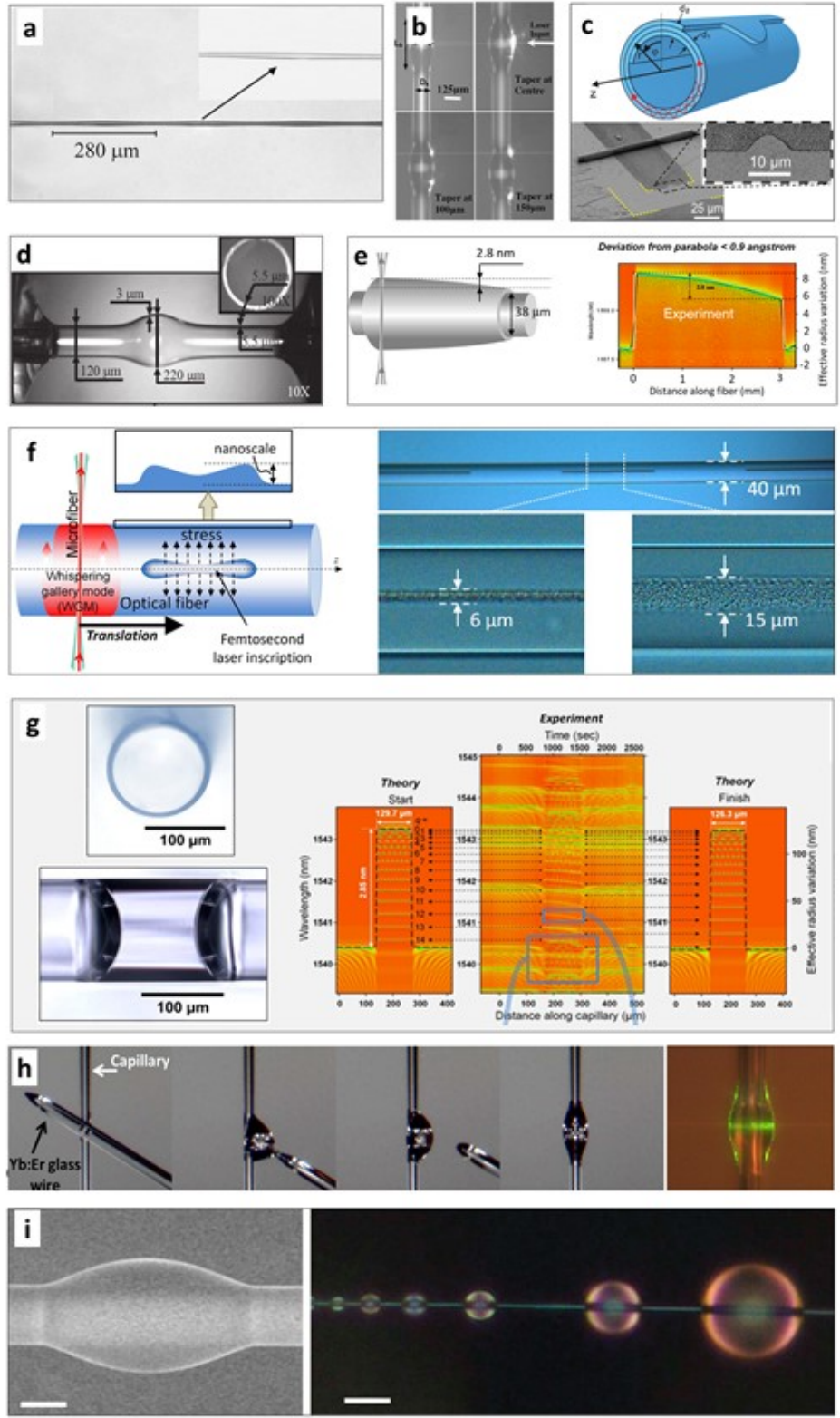

Fig. 8. Methods of BMR fabrication. (a) – Stretching a melted fiber [18]. (b) – Processing optical fiber in the fiber splicer [39]. (c) – Rolling of semiconductor bilayers [38]. (d) – Softening and pressurizing a microcapillary [42]. (e) – Fabrication of bottle microresonators by local annealing of an optical fiber [24]. (f) – Femtosecond laser inscription [49]. (g) – A BMR induced by a droplet in the microcapillary [28]. (h) – Depositing glass with lower melting temperature [47]. (i) – Depositing a polymer droplet solidified by heating [64]. (Reproduced with permission from Refs. [18, 24, 28, 38, 39, 47, 49, 64]).

### 3.4. SNAP technology

A super-precise method for fabrication of nanometer shallow BMR was developed in SNAP technology [20-28].The first demonstration of SNAP BMRs fabricated with angstrom precision was published in 2011 [22]. Originally, two fabrication methods were developed. The first method was based on modification of the effective radius variation of the optical fiber (including its physical size and refractive index variation) by local annealing with a $CO_2$ laser [22-26]. It was found that a few second local annealing of the optical fiber allows one the release the tension which was accumulated during fiber drawing. The release of tension causes a nanoscale variation in the fiber effective radius which is sufficient for the creation of BMRs. Depending on the power and duration of the exposure, it is possible to reproducibly fabricate BMRs with sub-angstrom precision [23-26]. As an example, Fig. 8(e) shows the illustration and spectrogram of resonant transmission spectrum of the BMR which was fabricated and demonstrated as the record low loss and small optical delay line in Ref. [24]. The second method of fabrication of SNAP BMRs is based on modification of the effective radius variation by local UV laser exposure which is applicable to fibers fabricated of photosensitive materials [22, 55]. Several other methods of fabrication of SNAP BMRs were developed recently. In Refs. [49, 53] the subangstrom precise fabrication of BMRs was demonstrated with a femtosecond laser. It was shown that the inscriptions introduced by the femtosecond laser internally pressurize the fiber and cause its nanoscale effective radius variation. In Ref. [52], parabolic BMRs were created using the $CO_2$ laser brushing technique for biconical taper fabrication. In Ref. [28] it was shown that BMRs can be induced by a droplet positioned inside a microcapillary with thin walls.

### 3.5. Polymer and soft glass BMRs

The authors of Ref. [46] developed a method for the fabrication of BMR using a UV-curable adhesive. The fabrication process included creating of liquid bottle-like microcavities along the waist of an optical fiber taper and solidifying it the liquids by UV light radiation. In Ref. [47], the authors describe a method for making BMR lasers by using a $CO_2$ laser radiation to melt Er:Yb glass positioned on a silica microcapillary or fiber (Fig. 8(h)). This method is based on the fact that the substrate silica glass has a higher melting point than the deposited glass. In Ref. [64], the polymer BMRs were fabricated using a self-assembly procedure (Fig. 8(i)). A silica fiber taper was used to deposit a microdroplet of R6G-doped epoxy resin solution on to the silica microfiber. After the deposition, the droplet shrinks due to the surface tension and formed a BMR which was solidified by heating.

### 3.6. Tunable and reconfigurable BMRs

Of special interest is fabrication of tunable BMRs. In Ref. [40, 42] the spectrum of a BMR was tuned by its mechanical stretching by pulling the ends of the fiber. It was demonstrated in Ref. [42] that the tuning range can be enhanced for hollow BMR (Fig. 8(d)). In Ref. [57], a temporal SNAP BMR introduced by local heating with a $CO_2$ laser was demonstrated. This microresonator was created, translated along the optical fiber with sub-angstrom precision in effective radius variation, and finally annihilated. In Ref. [51], a temporal SNAP BMR was introduced by local heating of a capillary fiber by a nonuniform wire positioned inside it. In particular, fine relative tuning of eigenfrequencies of two coupled SNAP BMRs was demonstrated.

### 3.7. Coupled BMRs and iterative fabrication method of SNAP structures

Before the invention of the SNAP technology, the fabrication of sequences of BMRs introduced along an optical fiber and coupled to each other was unfeasible. Furthermore, fabrication of coupled toroidal microresonators at the optical fiber surface, which have much smaller dimensions along the fiber axis as compared to BMRs and, thus, can be positioned much closer to each other (see [101, 102] for their theoretical description), is a challenging problem which has not been solved to date. The first demonstration of coupled SNAP BMRs was published in 2012 [44]. The SNAP fabrication technique was further developed in Refs. [25, 56] where the sub-angstrom fabrication precision was demonstrated.

The reason why the WGMs localized in SNAP BMRs introduced at the optical fiber surface can couple to each other is explained by the large value of their characteristic axial wavelength $\sim 100\ \mu m$ and small propagation constant $\sim 0.1\ \mu m^{-1}$ (see Section 2.2). As the result, WGMs can tunnel through the regions separating BMRs which may have the characteristic axial length of $100\ \mu m$ and more. This fact dramatically simplifies the fabrication of coupled BMRs.

In order to improve the fabrication precision of SNAP structures and, in particular, coupled BMRs, their iterative fabrication method has been developed [56]. This method consists of two steps illustrated in Fig. 9. The first step includes the exposure of the SNAP structure (Fig. 9(a)). At this step, the nanoscale variation of the effective optical fiber radius is introduced with a focused $CO_2$ laser beam by local annealing of the fiber. After the exposure is completed, the introduced structure is characterized by the microfiber scanning method [103, 104]. The characterization process illustrated in Fig 9(b) is the second step of the SNAP structure fabrication. During this step, the microfiber is scanned along the fiber under test touching it at points along the fiber axis where the spectra the introduced structure are measured. The spacing between the contact points (typically set to 1-20 $\mu m$) determines the axial resolution of measurements. The measured spectra collected in a SNAP spectrogram (see e.g., examples in Figs. 8(e) and (g) and Fig. 10) are used to determine if the fabricated structure fits the design specs or another iteration is required. In the latter case, the $CO_2$ power required to adjust for the measured deviations is calculated using the calibration method described in Ref. [56]. In brief, first determine the contribution of a single laser shot (typically several millisecond duration) to the variation of the effective fiber radius is measured. Next the structure is moved back to the exposure section of the setup to introduce the appropriate number of these shots to compensate for the measured deviation (Fig. 9(a)). The process is continued until the fabricated structure satisfies the required specifications.

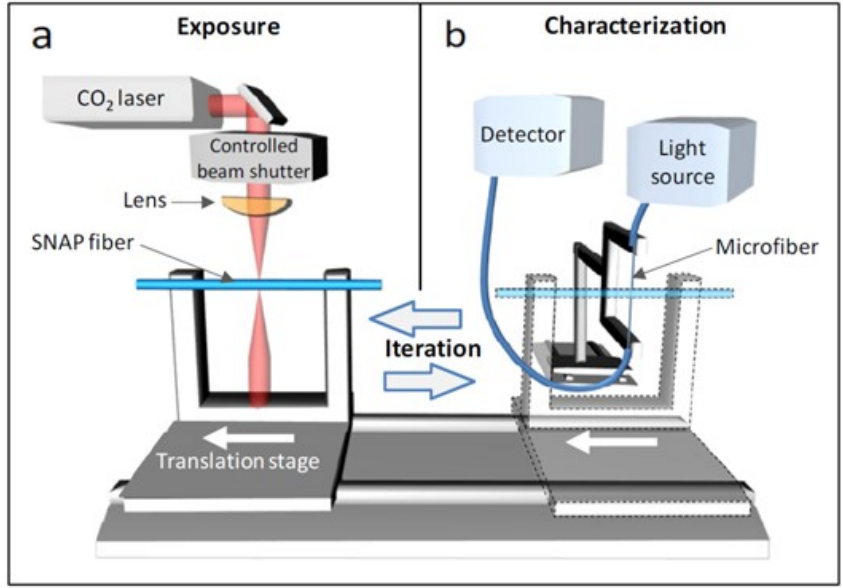

Fig. 9. Illustration of the setup for fabrication of SNAP devices. (a) – Exposure part of the setup; (b) – Characterization part of the setup. (Reproduced with permission from Ref [56]).

As an example of the iterative fabrication of couple BMRs is given in Fig. 10. In this experiment, 30 strongly coupled BMRs are created at the surface of a 19 μm radius fiber. The not-to-scale illustration of this structure is given in the inset of Fig. 10. Each of 30 BMRs spaced by 50 μm is introduced by a sequence of $CO_2$ laser exposures. Figure 10(a) shows the spectrogram of the introduced structure obtained with the microfiber scanning method described above. The spectrogram possesses the fundamental transmission band followed by a bandgap. The nonuniformity of the introduced structure measure from Fig. 10(a) was around 6 Å in effective radius variation. To correct this nonuniformity, the contribution of the laser exposure was first calibrated by introduction of linearly growing number of short $CO_2$ laser exposures along the same structure. The result of this introduction is shown in Fig. 10(b). The effect of an individual laser shot was then calculated by numerical comparison of spectrograms in Figs. 10(a) and 10(b). Finally, in order to make the resonators equal in effective radius, each of them was corrected by the appropriate number of laser shots. The result of correction is shown in Fig. 10(c). Numerical analysis of this spectrogram showed that the radius of the largest BMR was larger than the radius of the smallest BMR by the dramatically small value of less than 0.6 Å. The theoretical modelling of the introduced coupled BMR structure (Fig. 10(d)), which was in remarkable agreement with measured spectrogram (Fig. 10(c)), showed that the effective radius variation for each of the introduced BMR was approximately 2 nm.

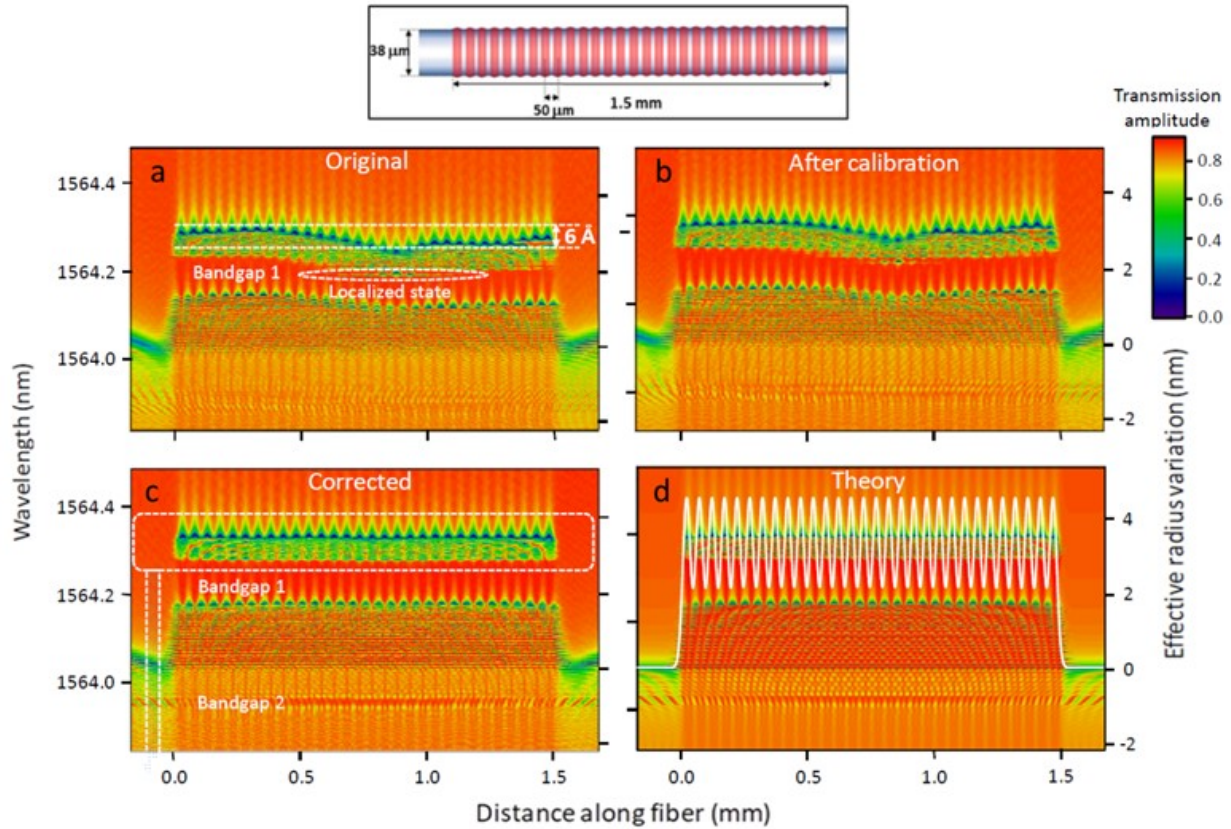

Fig. 10. Experimental and theoretical spectrograms of 30 coupled BMRs fabricated by iterations. The experimental spectrograms were obtained with 10 μm resolution along the fiber axis. (a) – The spectrogram of the originally fabricated coupled BMR structure. (b) – The spectrogram of this structure after calibration exposure. (c) – The spectrogram of the corrected structure. (d) – Theoretically modelled spectrogram of the fabricated structure. Inset – illustration of the fabricated 30 coupled BMRs (not to scale). (Reproduced with permission from Refs [56]).

### 3.8. Characterization of BMRs

Characterization of BMRs is usually performed with a microfiber scanned along its length. This method was originally demonstrated in Ref. [103] and developed in Ref. [104]. The basic idea of this method consists in the fact that due to very loss of optical fibers the shift of WGM resonances in the process of translation of the microfiber along the BMR under test can characterize its effective radius variation very accurately. The characterization precision demonstrated in Ref. [104] achieved 0.1 angstrom and can be further improved by more precise measurement of the WGM spectrum. The spectrograms obtained with this method are shown in Figs. 8(e) and (g). Even more accurate characterization of SNAP microresonators can be performed using a reference fiber method [105].

## 4. Applications of BMR

## 4.1. BMR delay lines

The BMR delay line is an alternative to miniature optical delay lines based on planar photonic technologies and, in particular, silicon photonics [106-111]. Prior to the demonstration of the BMR delay line in Ref. [24], the engineered miniature optical delay lines were fabricated of coupled ring resonators [106, 110] and coupled photonic crystal microcavities [109]. As compared to the latter structures, a BMR can act a fundamentally different type of delay line, which, in contrast to those proposed previously, is not based on the miniature photonic structures created by modulation of the material refractive index. Instead, as discussed in Sections 2.2-2.4 of this review, the slow propagation of light along the BMR axis is ensured by its rotation along the surface of an ultralow loss optical fiber. Remarkably, it was shown [24] that a BMR with semi-parabolic nanoscale effective radius variation can be impedance matched to the input-output microfiber (Fig. 11(a)) and perform the delay of 100 ps telecommunication pulses by several nanoseconds.

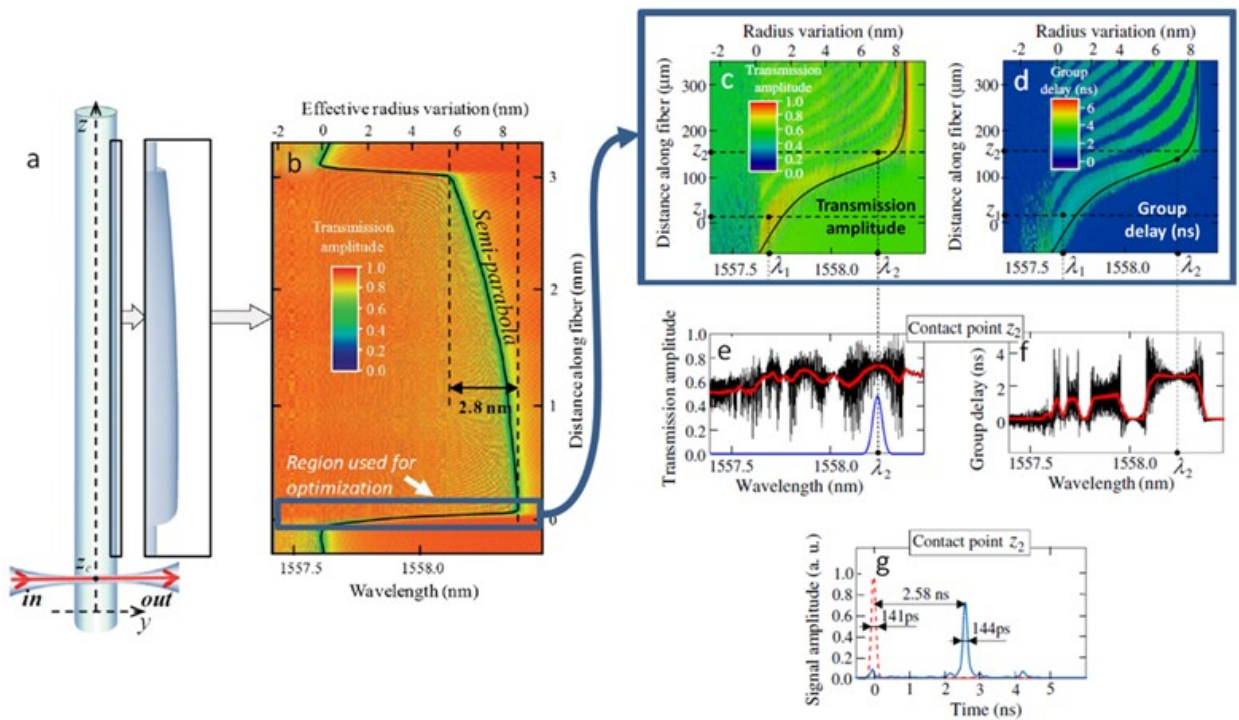

Fig. 11. (a) – Illustration of an BMR delay line. Inset – the magnified profile of the BMR radius variation. (b) – Experimental spectrogram of the fabricated BMR measured with resolution of 10 μm. (c), (d) – spectrograms of the transmission amplitude and group delay near the edge of the fabricated BMR measured after the optimization of coupling parameters. (e), (f) – Transmission amplitude and group delay spectra at the point $z_2$ along the BMR axis corresponding to minimum spectral oscillations in spectrograms (c) and (d), respectively. (g) – The output pulse profile (blue solid line) calculated for the input 100 ps pulse (red dashed line) from the spectrum shown in (e) and (f). (Reproduced with permission from Refs [24]).

Here we briefly describe the BMR delay line demonstrated in Ref. [24], which dramatically surpasses previous designs both in the achieved small internal losses and miniature dimensions. The 3 mm long BMR was created at the 19 μm radius optical fiber by annealing with a $CO_2$ laser beam (Fig. 11(a)). The speed of the beam was varied in the process of translation along the fiber to ensure the required nanoscale semi-parabolic effective radius variation. The depth of the introduced BMR was 8 nm, while its semi-parabolic part, which was introduced with a precision of better than 0.9 Å, had the depth of 2.8 nm (Fig. 11(b)).

The coupling between the input-out microfiber and BMR was optimized in Ref. [24] by translating the microfiber along its axis and along the BMR axis in the region outlined in Fig. 11(b) by a blue bold rectangle situated near the BMR edge. The spectrograms of the transmission amplitude and group delays in this rectangle measured with the optimized microfiber coupling is shown in Fig. 11(c) and Fig. 11(d). As an example, Fig. 11(g) shows the profiles of the input 100 ps pulse and of the corresponding output pulse found from the spectra shown in Fig. 11(e) (transmission amplitude) and Fig. 11(f) (group delay) at optimized microfiber position $z_2$. Fig. 11(g) demonstrates the pulse delay of 2.58 ns with intrinsic loss of only 0.44 dB/ns and no dispersion.

The described BMR delay line presents a solution of one of the central problem of photonics – creation of a delay line having the record small dimensions and insertion losses which is required for several emerging engineering applications of modern photonics including optical signal processing at microscale. Worth noting that the BMR device presented in this section can be further miniaturized if it designed and fabricated of SNAP coupled BMRs as proposed in [60]. The future development of the described BMR delay line paves the way to the creation of realistic miniature optical buffers [31]. BMR with different effective radius variation (other than parabolic) can be designed for the required transformation of optical pulses. For example, the BMR which effective radius variation depends on the resonator length $z$ as $\sim z^{2/3}$ can be used as a miniature dispersion compensator experimentally demonstrated in Ref. [59].

## 4.2. BMR lasers

Different types of lasers with microscopic dimensions have been proposed and demonstrated. Usually, they are based on the optical microresonators fabricated of active materials such as ring, spherical, toroidal, capillary, and bottle microresonators. A review of microscopic lasers based on WGM microresonators proposed and demonstrated up to 2006 can be found in Refs. [4, 5] and a later progress in the field is reviewed in Ref. [10]. While these lasers have different geometry, the physical principles of their operation are similar. In particular, due to the high Q-factor of microresonators, these lasers have a small lasing threshold. Lasing microresonators can be fabricated of an active material or fabricated of a passive material and post-processed by coating with an active material.

The BMR lasers [61-68] have certain advantages compared to other microlasers. In fact, their elongated geometry and fabrication simplicity allows one to fabricate active BMR with predetermined spectrum and facilitate mode selection by patterned pumping much easier than for active microresonators having other geometries. Several approached for fabrication of BMR lasers were developed. In Ref. [61], low threshold high Q-factor BMR Raman laser was fabricated of a silica capillary. In Refs. [62, 64, 65, 67, 68], active BMRs were fabricated by depositing polymer droplets doped with active materials onto the optical fiber followed by the UV and/or heat curing. In Ref. [63] BMR lasers were fabricated by depositing Er:Yb doped glass, which was melted by the $CO_2$ laser, onto silica microcapillaries and fibers. The deposition was possible since the doped glass had lower melting temperature than that of silica. In Ref. [65] Brillouin lasing and Brillouin-coupled four-wave-mixing in an ultra-high-Q silica BMR was demonstrated. In Ref. [72] design of and rare-earth-doped BMR lasers, which were optimized for a low threshold pump power, high efficiency and predetermined lasing wavelength, was performed.

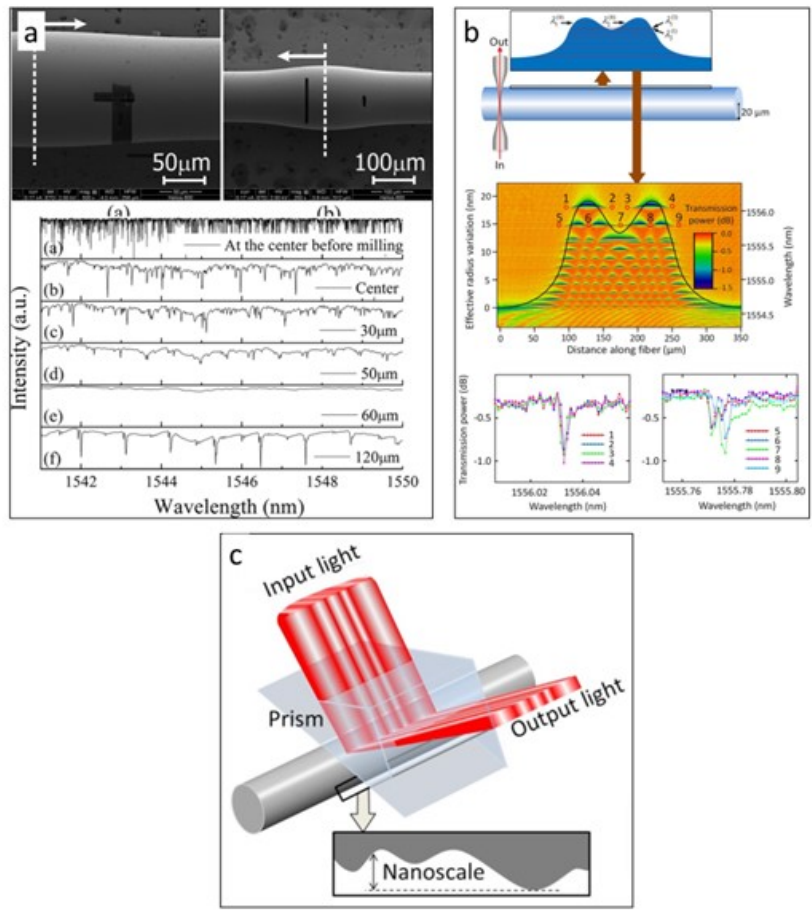

Fig. 12. (a) – Spectral cleaning of bottle microresonators [50]. (b) – Fabrication of coupled SNAP BMR with subangstrom precision [26]. (c) – Combination of the spatial pump beam engineering, ultraprecise SNAP platform, and spectral cleaning for fabrication of BMR lasers with required characteristics [19]. (Reproduced with permission from Refs [19, 26, 50]).

It is of great interest to discuss the advantages of BMR lasers as compared to other types of miniature lasers. These advantages are primarily caused by the characteristic elongated geometry of BMRs which allows for the design and accurate fabrication of BMR lasers with predetermined spectrum. Three basic methods of the BMR laser design and fabrication, which can be used separately or in combination with each other, are known.

The first method of the BMR laser design and fabrication can be applied if the light spectrum emitted by the BMR laser has to be cleaned from the unwanted peaks. If the BMR modes corresponding to these peaks are known, then these peaks can be removed by destroying the related modes. To this end, the authors of Ref. [50] demonstrated a method of spectral cleaning by inscribing microgroove scars on the BMR surface using the focused ion beam milling method (Fig. 12(a)). Obviously, in order to destroy the unwanted mode and keep the required mode (or modes) untouched, the introduced grooves should be as narrow as possible and positioned along the line corresponding to nodes of all the required modes.

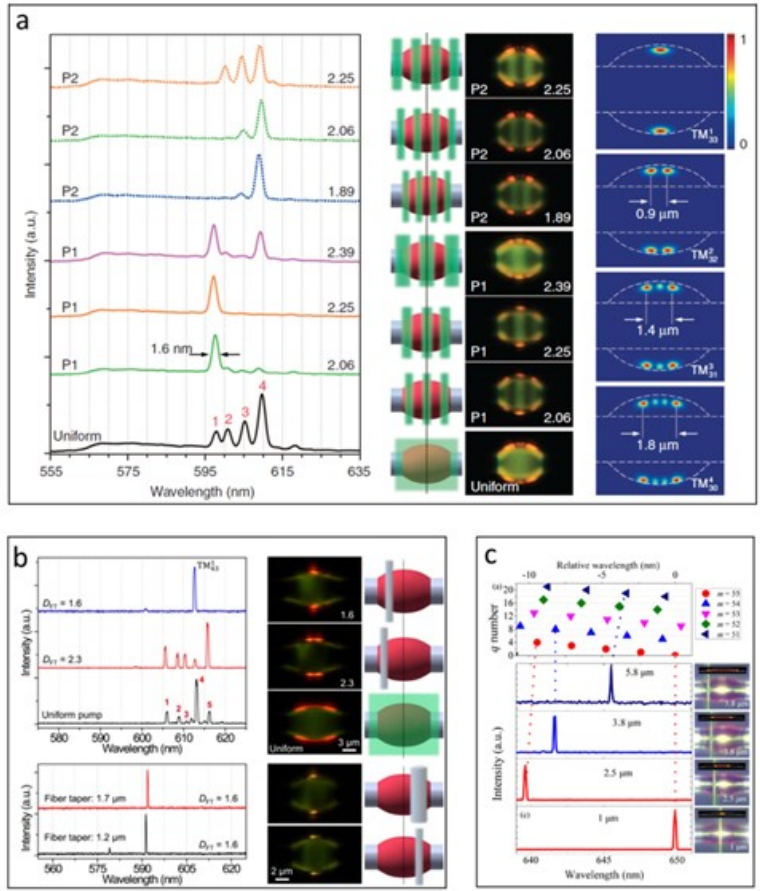

Fig. 13. WGM lasing in polymer BMRs pumped with an external structured laser beam and radiation launched by a coupled microfiber. (a) – Lasing spectra and corresponding microscope images and electric field-intensity distributions of a polymer BMR under the action of a uniform and a spatially modulated laser beam [64]. (b) – Lasing spectra and corresponding microscope images of a polymer BMR pumped by the uniform radiation and by the radiation from a microfiber contacting BMR at different positions along the BMR axis [65]. (c) – Single mode lasing at different frequencies for different positions of a coupled microfiber along the BMRs and the corresponding optical microscope images of the lasing mode field distribution [67]. (Reproduced with permission from Refs [64, 65, 67]).

The second method of the BMR laser design and fabrication is based on the SNAP technology, which allows for ultraprecise fabrication of lasing BMR having the predetermined effective radius variation. As an example, Fig. 12(b) shows two coupled SNAP BMR [26], which, if fabricated of an active material, can be designed to lase at two frequencies with very small separation. The BMRs are fabricated to be similar in size so that the splitting of their eigenfrequencies exponentially decreases with the separation between the BMRs. In Fig. 12(b), while the splitting between the fundamental modes of these resonators with axial quantum number $q = 0$ cannot be observed, the splitting between modes with the axial quantum number $q = 1$ is detected [26]. Notice that the modes with $q = 0$ and $2$ adjacent to that with $q = 1$ of our interest can be removed with the spectral cleaning method. To this end, the grooves destroying these modes should be placed in the middle of each of the coupled BMRs corresponding to the nodes of the $q = 1$ mode.

The third method of the design and fabrication of BMR lasers, which was proposed and demonstrated in Ref. [64], is concerned with the spatial engineering of the input pump beam (Fig. 13(a)). This method, as well as the first two, explores the spatial elongation of BMRs, which can be used to appropriately distribute pumping along the resonator length. It is seen from Fig. 13(a) that depending on the distribution of the pumping radiation along the BMR, the latter can lase at certain single or multiple frequencies. In other approaches [65, 67], local pumping of BMR was performed with a coupled microfiber. In this case, again, depending on the position of the microfiber, the BMR can lase at a single of multiple frequencies (Fig. 13(b) and (c)).

Fig. 12(c) illustrates a device where all the described methods of design and fabrication of BMR lasers can be combined in the most effective way [19]. The active BMR is supposed to be designed and fabricated with the ultrahigh precision using the SNAP technology. The input pump beam is delivered to the BMR evanescently through a prism. As compared to the free space beam pumping, this allows to minimize the power consumption due to the resonant amplification of the pump intensity in the BMR. Typically, the gap between the prism and BMR, which ensures the effective evanescent coupling is of the order of 100 nm. Remarkably, the required nanoscale variation of the effective BMR radius along its length fits in the gap between the prism and BMR.

## 4.3. Nonlinear BMRs

Nonlinear photonics is one of the fastest emerging branches of modern photonics. Its several most exciting recent applications were demonstrated for the high Q-factor optical microresonators where the electromagnetic field can be dramatically enhanced leading to strong nonlinear effects. The theory and applications of non-linear processes in high Q-factor optical microresonators, in particular, in WGM microresonators, have been intensively developed over the last decades (see e.g., Ref. [9] for the review). What is advantageously special in nonlinear phenomena in BMRs as compared to other types of microresonators? The answer lies again in the benefits of the elongated geometry of BMRs and possibility of their ultra-precise fabrication based on the SNAP technology.

An interesting example of a nonlinear BMR device is an all-optical switch based on the Kerr effect [69, 71]. If the BMR material exhibits the third-order susceptibility, its refractive index depends on the WGM intensity $I$ as $n = n_r + n_2 I$ where $n_r$ is the refractive index of the resonator material and $n_2$ is its Kerr nonlinear refractive index. Generally, the shift of the resonance frequency is proportional to $n_2 Q^2 / V$ where $Q$ and $V$ is the WGM Q-factor and its volume. Though the value o $n_2$ is very small, the nonlinear effect can be strong even for the relatively small input power provided that the BMR modes have sufficiently high Q-factor and small volume. The authors of Ref. [69, 71] demonstrated a BMR which had one of the highest value of $Q^2 / V$ for optical microresonators. As the result, they observed the bistable BMR behavior caused by the nonlinear Kerr effect at very low powers and demonstrated a single-wavelength all-optical switch at a record-low threshold of 50 mW. In this experiment, illustrated in Fig. 14(a), the signal light was repeatedly switched between the bus and the drop microfibers with a rate of 1 MHz.

In another experiment [74], a hollow BMR was fabricated from a microcapillary (inset in Fig. 14(b)). In order to excite a WGM, the input-output microfiber taper was positioned at approximately 25 μm from the BMR center. At this position, the authors of [74] observed two sideband peaks of equal height in the vicinity of the pump wavelength (Fig. 14(b)) which arise due to the four wave mixing. In order to avoid the excitation of Raman transitions, the pump power was chosen in the range from 3.3 mW to 4.2 mW, which is above the four wave mixing threshold and below the Raman threshold for this WGM.

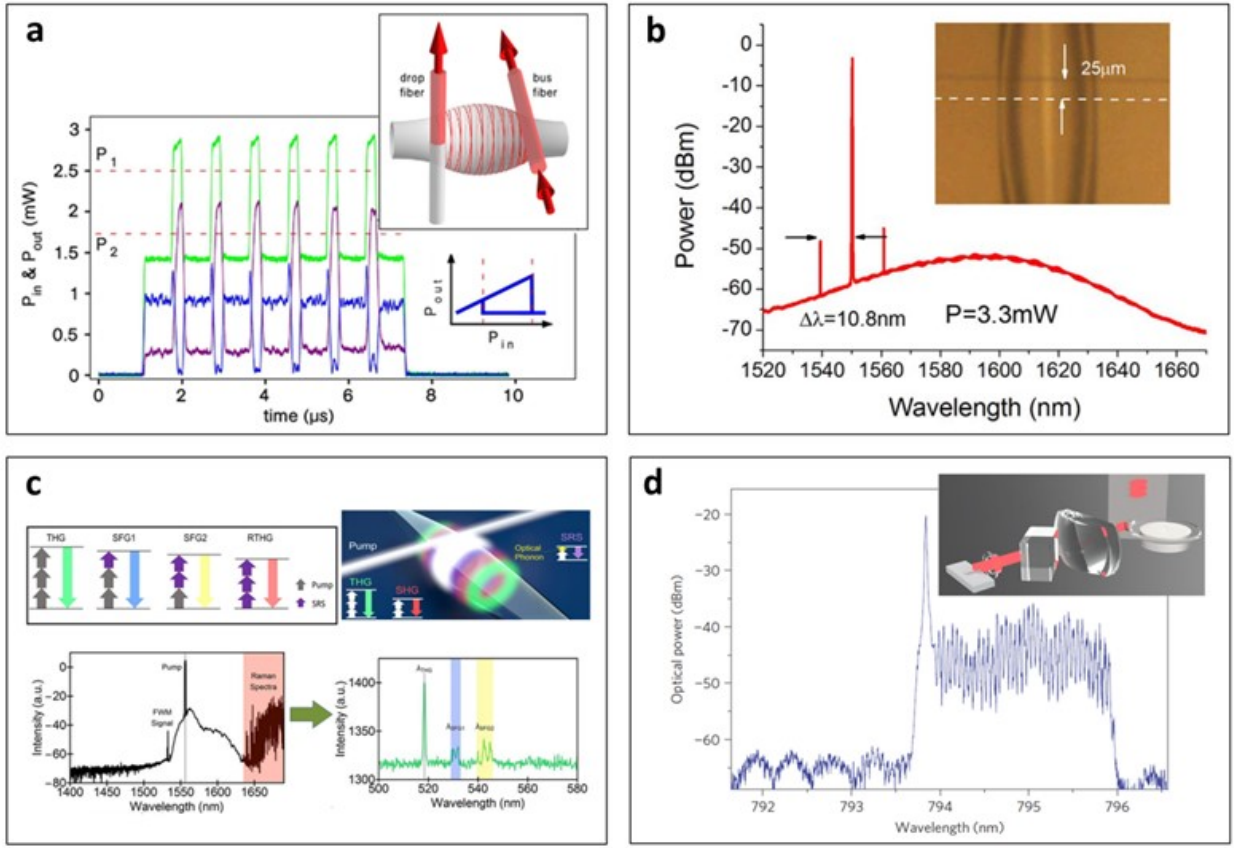

Fig. 14. (a) – A BMR all-optical switch demonstrated in Ref. [69]. The BMR is coupled to the drop and bus microfibers illustrated in the inset. The dashed lines in the small inset plot correspond to the two levels which are located below and above the bistable regime. (b) – The transmission spectrum of the BMR shown in the inset demonstrating the four-wave mixing [74]. (c) – Illustration of the BMR (upper right inset), the energy diagrams of nonlinear processes (upper left inset) and observed experimental spectra from Ref. [73]. (d) – The spectrum of frequency comb generated along the axial WGM series in Ref [70]. Inset shows the experimental setup including the toroidal resonator. (Reproduced with permission from Refs [69, 70, 73, 74]).

The authors of Ref. [72] observed nonlinear processes in BMRs which included the third harmonic generation via the four wave mixing and the second harmonic generation via the three wave mixing processes. The BMR (illustrated in the right inset of Fig. 14(c)) had the diameter of 64.5 μm in the center and two necks separated by 3.5 mm away from each and diameter of 23.1 μm. The energy diagrams of nonlinear processes observed in Ref. [72] are shown in the left inset of Fig. 14(c). They include the sum frequency generation by mixing of two pumps and one Raman excitation (SFG1), one pump and two Raman excitations (SFG2), and third harmonic generation from Raman light generated by stimulated Raman scattering (RTHG). The spectra showing the observed nonlinear processes are shown in Fig. 14(c). The wavelength and the power of the pump light were 1556.1 nm and 208.3 mW (Fig. 14(c), left). The observed optical spectra of visible light generated by these processes are shown in the right plot in Fig. 14(c).

Another example of an experimentally observed nonlinear process in an optical microresonator reported in Ref. [70] (Fig. 14(d)) has an important relation to a similar effect in a BMR. Commonly, optical resonators with normal dispersion (i.e., where the free spectral range grows with wavelength) exhibit a much weaker four wave mixing process, which is difficult to observe since it is often comparable with the effect of stimulated Raman scattering. The free spectral range of a WGM microresonator with radius $r$ along the azimuthal quantum number $m >> 1$ is $\Delta \lambda_{az} \approx \lambda^2 / (2\pi n_r r)$ and, thus, corresponds to the normal geometric dispersion. Therefore, to arrive at the anomalous dispersion, the resonator should be fabricated of a material which anomalous dispersion which can compensate the normal geometric dispersion. For a silica microresonator, this fact significantly complicates the generation of frequency combs for the pump wavelength smaller than 1.3 μm. However, the shape of a BMR can be designed to arrive at the required geometric dispersion of WGMs along the axial quantum number $q$, which can be found, e.g., from the semiclassical quantization rule of Eq.(1). The authors of Ref. [70] reported the first demonstration of the frequency comb generation along the axial quantum number for the toroidal resonator for the arbitrary pump frequency. The resonator used in Ref. [70] and a sample frequency comb spectrum measured are shown in Fig. 14(d).

The experimental results [70] suggest that the optical frequency comb with the predetermined repetition rate and central frequency can be generated by the appropriately designed BMR. The numerical simulations of BMR frequency comb generators [32-35] reviewed in Section 2.6 support this prediction. The SNAP technology, which allows for the ultraprecise fabrication of such microresonators, makes their demonstration feasible in the nearest future. Remarkably, since the axial radius of BMRs can be dramatically large [24], the repetition rate of BMR frequency comb generators can be very small. In Ref. [33] it was shown that BMRs are capable to generate frequency combs which have small repetition rate and broad band simultaneously. The idea of Ref. [33] is illustrated in Fig. 15. The inset of Fig. 15(a) shows a BMR with cosine-shaped radius variation which corresponds to the constant axial free spectral range (see Eqs. (2) and (3)). It was suggested in [33] that if the BMR azimuthal free spectral range is the exact multiple of its axial free spectral range (Fig. 15(a)) then this resonator can be designed to ensure the nonlinear excitation of frequency combs due to transition both along the axial and azimuthal resonance series (Fig. 15(b)). Alternatively, a narrow band axial resonance series situated between adjacent azimuthal resonances can be excited by pumping with a mode-locked laser followed by excitation of similar series corresponding to other azimuthal quantum numbers (Fig. 15(c)).

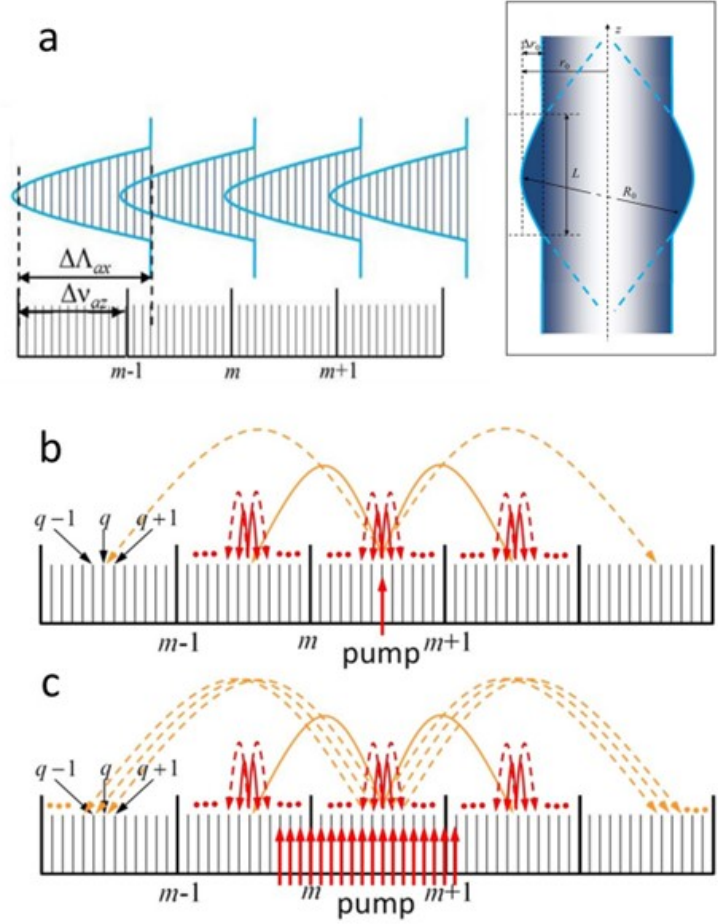

Fig. 15. (a) – Illustration of the BMR spectrum having the azimuthal free spectral range equal to the exact multiple of axial free spectral range. Inset shows a BMR with cosine-shaped radius variation. (b) – The broadband frequency comb generation with a single-frequency laser pump. (c) – The broadband frequency comb generation with a narrowband comb pump generated by a mode-locked laser. (Reproduced with permission from Ref. [33]).

## 4.4. Optomechanics of BMRs

Optomechanical phenomena in BMRs are in many cases similar to those in other optical microresonators [112]. The major differences are, again, caused by the elongated geometry of BMRs. Due to the interaction of an optical WGM of the resonator having frequency $\omega_r^{(opt)}$ and its mechanical mode having frequency $\omega^{(mech)}$, the resonant optical frequency can split into series $\omega_r^{(opt)} \pm m\omega^{(mech)}$ where $m$ is a positive integer and $\pm$ correspond to the Stokes and anti-Stokes transitions. In the description of the excitation of these series illustrated in Fig. 13(a) we follow Ref. [77]. The setup used in Ref [77] is shown in Fig. 13(b). The pump light, which was generated by a tunable laser with a linewidth of 300 kHz in the 1550 nm band and amplified, was controlled by a polarization controller (PC) and an attenuator (Att.). The pump light was coupled into the BMR with an input-output tapered fiber and transmitted through a photodetector (PD) to a digital storage oscilloscope (DSO) and an electrical spectrum analyzer (ESA) used for the measurement of spectrum of the BMR mechanical oscillations. As an example, Figs. 16(c1)-(c4) show the measured spectra of mechanical oscillations for the BMR with characteristic dimensions 100μm × 30μm × 4mm for the pump powers of (c1) 24.1 mW, (c2) 22.8 mW, (c3) 20.8 mW, and (c4) 20.1 mW. The excited mechanical frequency was $\omega^{(mech)} = 2\pi \cdot 35.6$ MHz. At a pump power of 24.1 mW, the authors of Ref. [77] observed up to 13 equally spaced resonances (harmonics). The number of resonances gradually decreased (i.e., the higher harmonics of mechanical oscillations gradually disappeared) with decreasing the pump power (Figs. 16(c1)–(c4)).

An interesting application of BMRs was demonstrated in Refs. [82, 83] where the optomechanics of capillary BMRs filled with liquid was investigated. The authors demonstrated a microfluidic system which was actuated optomechanically, thereby bridging microfluidics and optomechanics. The hollow BMR used in Ref. [83] is shown in Fig. 16(d). Its characteristic dimensions were 100μm × 35μm × 1.1mm. The setup used for the optical excitation of mechanical vibrations is illustrated in Fig. 16(e). Light was coupled in and out of the BMR with a tapered optical fiber which was not in direct contact with the resonator. The spectra of experimentally observed mechanical vibrations of the BMR filled with water are shown in Fig. 16(f1)-(f4). The measured mechanical frequencies ranged from 99 MHz to 11 GHz. The Q-factor of the BMR exceeded $10^8$. Overall, it was experimentally demonstrated that the optomechanical interactions in the capillary BMR depend on the properties of fluid inside it. The authors suggested that the demonstrated high quality-factor mechanical modes may enable optomechanical interaction with chemical and biological species, which are simultaneously strongly localized and highly-sensitive and, thus, can be used as a new type of the microfluidic sensing device (see Section 4.6).

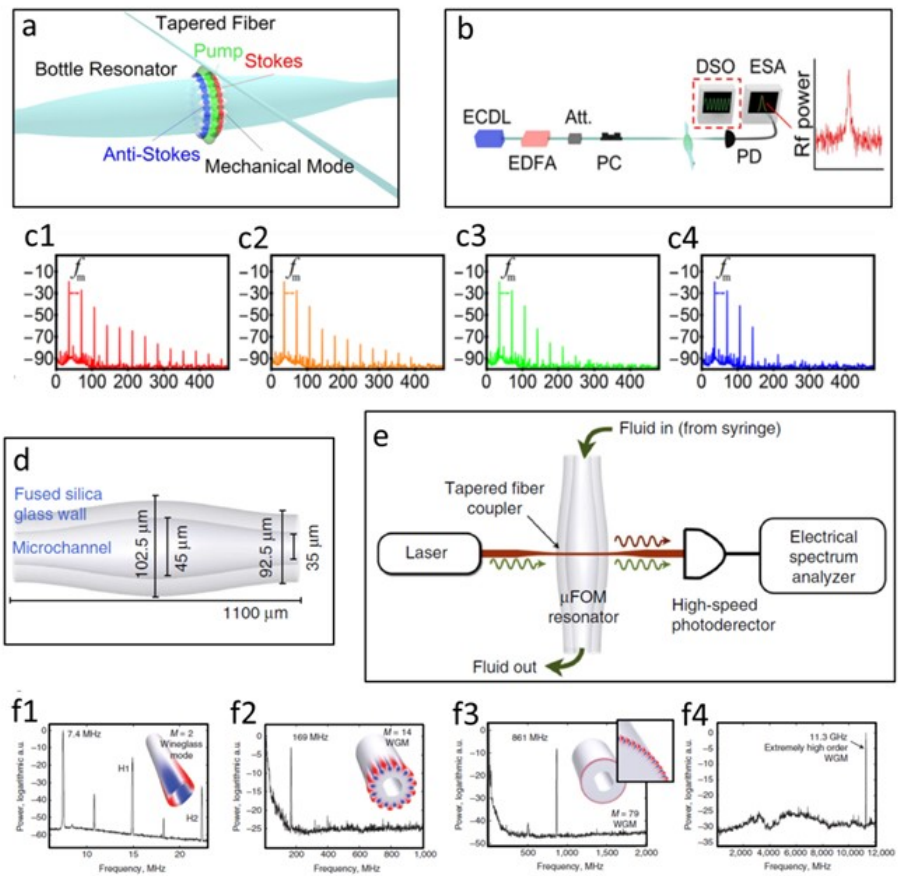

Fig. 16. (a) – The illustration of an optical BMR used for excitation of mechanical vibrations [77]. (b) – The illustration of the experimental setup for the measurement of the spectrum of mechanical vibrations generated in the BMR [77]. (c1)-(c4) – Experimental spectra of mechanical oscillations in a BMR observed in Ref. [77]. (d) – Dimensions of the capillary BMR used for excitation of mechanical vibrations in Ref. [83] (not to scale). (f1)-(f4) – Experimental spectra of mechanical oscillations in a water-filled BMR [83].(Reproduced with permission from Refs. [77, 83]).

Understanding the properties of mechanical modes as compared to the optical WGMs in elongated BMR is of major importance for the development of BMR optomechanics. This problem was addressed in Ref. [79] where it was shown that, in analogy to optical BMRs, there exist acoustic BMRs where modes can be localized by extremely small nanometer-scale radius variation. In addition, it was shown that, in contrast to optical BMRs, there exist acoustic antibottle microresonators having the shape of a neck rather than a bulge (Fig. 17(a)). The free spectral range of eigenfrequencies of the acoustic microresonators with parabolic radius variation is proportional to $(-\Lambda_n R r_0)^{-1/2}$ where $R$ is the axial bottle radius (negative for regular BMRs) and $r_0$ is the azimuthal radius. The dependence of parameter $\Lambda_n$ for acoustic modes $\mathrm{LP}_{0n}$ with azimuthal quantum number $m = 0$ and radial quantum number $n$ is shown in Fig. 17(b). It is seen that the parameter $\Lambda_n$ can be both positive (bottles) and negative (antibottles). Fig. 17(d) compares the axial distribution of acoustic and optical modes for the silica BMR with nanoscale parabolic radius variation with axial radius $R = 824$ m shown in Fig. 17(c). Remarkably, the characteristic length of the acoustic modes with the same axial quantum number is much greater than that for the optical modes.

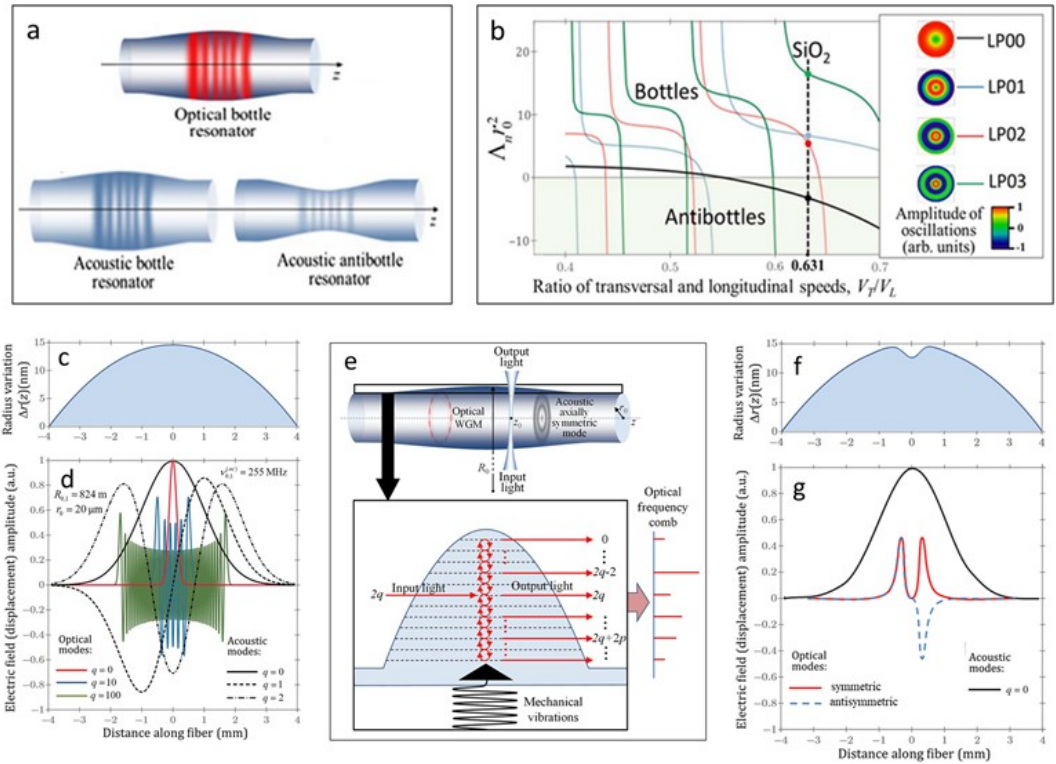

Fig. 17. (a) – Illustration of optical BMR, acoustic BMR, and acoustic anti-BMR. (b) – Dimensionless parameters $\Lambda_n r_0^2$ as a function of ratio between the transverse and longitudinal speeds of sound, $V_T / V_L$, of the BMR material for acoustic modes with azimuthal quantum number $m = 0$ and radial quantum numbers $n = 0, 1, 2$, and 3 (LP01, LP02, and LP03). Inset: cross-sectional distribution of amplitudes of modes LP01, LP02, and LP03 for a silica fiber. (c) – Nanoscale parabolic radius variation of the BMR. (d) – Distribution of the amplitudes of optical and acoustic modes along the BMR with the parabolic profile shown in (c). (e) – Illustration of the optical frequency comb generated mechanically. (f) – BMR radius variation with a double bump in the middle. (g) – Distribution of the amplitudes of optical symmetric (red) and antisymmetric (dashed-blue) and acoustic (black) modes along the BMR with the radius variation shown in (f). (Reproduced with permission from Refs. [78, 79]).

It was shown that the BMR with gigantic axial radius of the order of 1 km can be designed so that its mechanical eigenfrequency matches the separation of its optical eigenfrequencies along the axial quantum number [78, 79]. Since the axial free spectral range of acoustic modes is very small, it is desirable for this application to remove all acoustic modes except for the fundamental axial mode (black line in Fig. 17(c)). This can be done by appropriate modification of the profile of the bottle resonator outside the location of the fundamental acoustic mode. Excitation of an acoustic mode, which frequency matches the axial free spectral range of optical modes, makes possible the generation of optical frequency comb mechanically (Fig. 17(e)). In contrast to the series of resonant frequencies shown in Fig. 17 (c) and (f), these optical resonances are generated by the external excitation of mechanical vibrations rather than optical pumping [78].

Another way of creating of a BMR where the separation of optical WGM frequencies matches a mechanical eigenfrequency is illustrated in Fig. 17(g) and (f). The radius variation of this BMR has a double-bump profile in the middle (Fig. 17(g)) corresponding to two coupled BMRs which support two evanescently coupled (symmetric and antisymmetric) optical WGMs (red and dashed-blue lines in Fig. 17(f)). By adjusting the separation between the bumps, the optical mode splitting can be made equal to the eigenfrequency of the acoustic mode of this resonator (black line in Fig. 17(f)). Experimentally, a structure of two identical coupled BMRs fabricated with the precision better than 0.02 nm in effective radius variation was demonstrated in [26] (Fig. 12(b)).

## 4.5. BMRs for quantum processing

Cavity quantum electrodynamics (CQED) investigates quantum phenomena in the interaction between photons, atoms, and other particles or excitations in an optical cavity [109]. Several theoretical proposals of applications of BMR in CQED and experimental demonstrations of CQED phenomena in BMRs have been published [84-87].

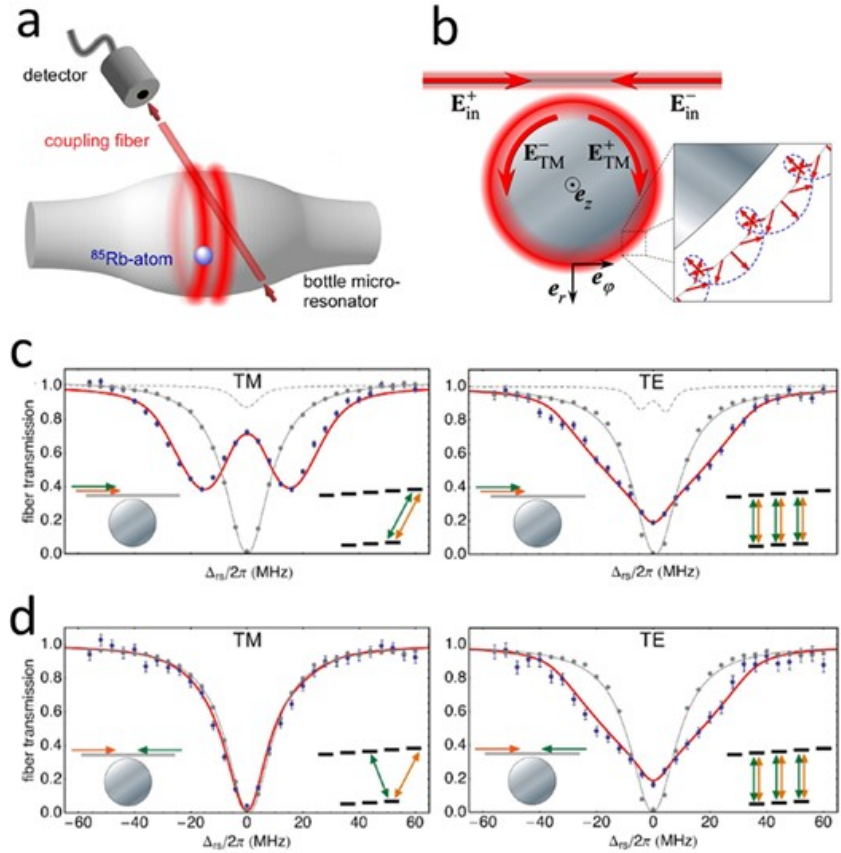

Fig. 18. (a) – Illustration of a BMR coupled to rubidium atom. (b) – Spatial dependence of the electric field for TM modes. (c) – Measured fiber transmission for coupling of an atom to TM mode. (d) – The same measurement for the TE mode. The insets in (c) and (d) show the direction of the detection light (orange) and the spectroscopy light (green) and the scheme of atomic level transitions. (Reproduced with permission from Ref. [85]).

As in all types of WGM microresonators, modes in BMRs can be separated into those having the electric field polarization, which either lies in the plane perpendicular to the BMR axis (TM) or is directed along this z axis (TE). Authors of Ref. [85], investigating $^{85}$Rb atoms strongly coupled to a BMR (Fig. 18(a)), recalled that while the TE modes are transversally polarized, the TM modes are non-transversally polarized (see e.g., Ref. [113] for detail explanation). This means that that the electric field of TM modes has a component along the azimuthal as well as the radial directions (Fig. 18(b)). In particular, the electric field vector of TE modes is perpendicular to their wave vector at any position of the mode. In contrast, the TM modes are nontransversally polarized, i.e., their electric field vector has a nonvanishing component along the wave vector. This longitudinal component oscillates with the 90⁰ out of phase with respect to the transversal component. In the experiment [85], the frequency of the BMR was tuned to an atomic transition frequency. The microfiber shown in Fig. 18(a) was critically coupled to the BMR. The polarization of the input resonant light was matched to the polarization of the BMR mode. As a result, the light was entirely coupled into the BMR mode and dissipated. Next, a cloud of laser-cooled $^{85}$Rb atoms was launched towards the BMR. If an atom is situated in the evanescent field of the BMR mode it strongly interacts with this mode. Then, the Rabi splitting of the resonance frequency [114] results in a detectable increase of the coupling fiber transmission. To measure the spectrum, a fast optical switch was used to turn off the detection light in real time and to simultaneously apply the spectroscopy light through the coupling fiber. Figs. 18(c) and (d) show the significant difference of transmission spectra of the TM and TE modes, which follows from the different behavior of the TE and TM modes indicated above, in a good agreement with theory [85].

A setup similar to that shown in Fig. 18(a), allowed the authors of Ref. [84] to demonstrate that an ultra-strong interaction of atoms and light can lead to a nonlinear $\pi$ phase shift for coincident photons. To experimentally determine this shift, the transmitted light was analyzed using three polarization bases by recording coincidence counts between the different detectors. It was suggested that the demonstrated conditional $\pi$ phase shift can be the basis for realizing a single-photon transistor and potentially enable deterministic quantum computation protocols with photons.

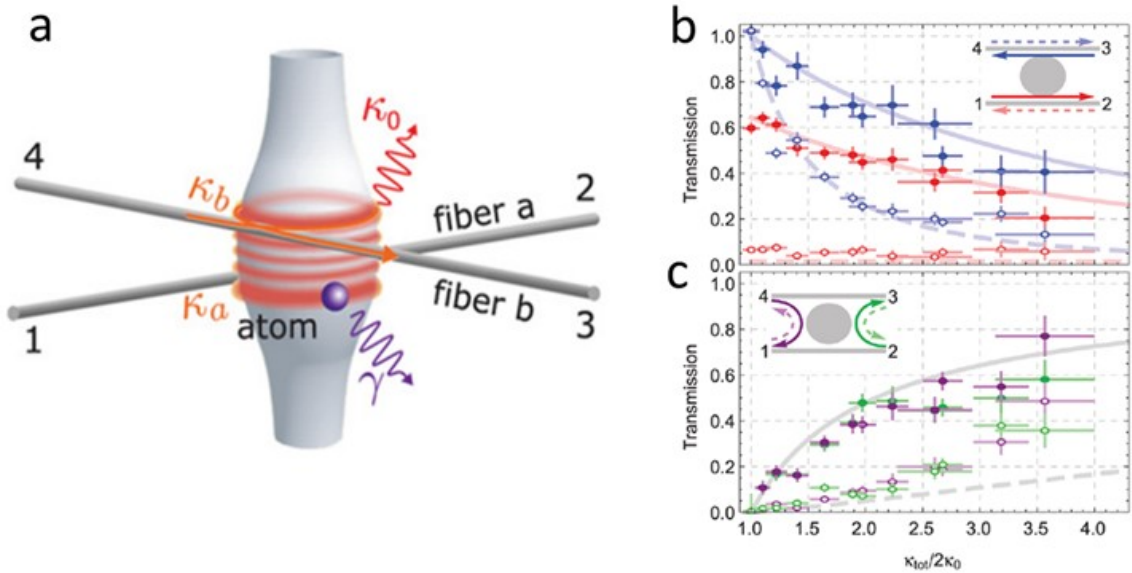

Fig. 19. (a) – Illustration of a BMR coupled to two input-output microfibers and a single $^{85}$Rb atom (b) and (c) – Measured port-to-port transmission efficiency as a function of the of the ratio $\kappa_{tot}/\kappa_0$ between the total field decay rate of the BMR in the presence of an atom $\kappa_{tot}$ and the intrinsic field decay rate of the BMR $\kappa_0 = 2\pi \cdot 5$ MHz. (Reproduced with permission from Ref. [87]).

Using a BMR coupled to two input-output microfiber tapers illustrated in Fig. 19(a), the authors of Ref. [87] demonstrated a fiber-integrated quantum optical circulator operated by a single $^{85}$Rb atom. In order to achieve efficient routing, the coupling to microfibers were tuned close to each other. It has been shown that, depending on the prepared internal state of the atom, this device can route light either from the input port $i$ to the adjacent output port $i + 1$ with $i \in \{1, 2, 3, 4\}$ (Fig. 19 (a)) or in the reversed direction. Figs. 19(b) and (c) show the measured port-to-port transmission efficiency as a function of the ratio between the total field decay rate of the BMR in the presence of an atom $\kappa_{tot}$ and the intrinsic field decay rate of the BMR $\kappa_0 = 2\pi \cdot 5$ MHz.

A system of coupled SNAP BMRs [25] illustrated in Fig. 20(a) has been proposed as a device which can be used as an efficient frequency converter of a single photon operated without external field pumping [88]. The distance between three coupled BMRs shown in this figure is chosen so that two WGMs, $|u_1\rangle$ and $|u_2\rangle$, with the frequencies close to the input and output frequencies, $\omega_0$ and $\omega_0 \pm \omega_a$, of a photon have the distributions illustrated in Fig. 20(b). The input photon enters the system by resonant transmission into mode $|u_1\rangle$, then experiences the resonant down or up conversion to the frequency of mode $|u_2\rangle$ due to interaction with atoms positioned in the central BMW, and finally exits the system with the acquired frequency $\omega_0 \pm \omega_a$. It was shown that, under certain conditions, the probability this process can approach unity [88].

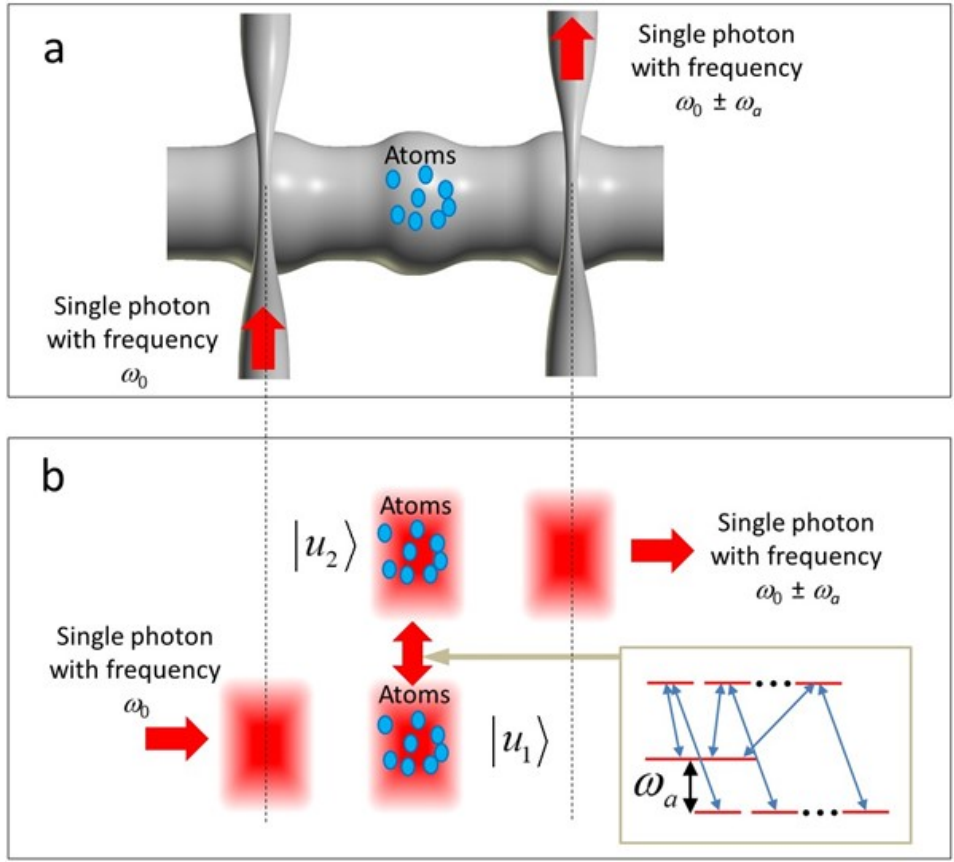

Fig. 20. (a) – Three coupled SNAP BMRs with the atomic cloud near the center resonator. (b) – Illustration of distribution of WGMs and atoms. Inset: three-level atomic transitions realizing the frequency conversion of a single photon.

## 4.6. BMR sensors

Similar to other WGM resonators, BMRs can be used for precise optical, mechanical, physical chemical, and biological sensing [89-97]. Remarkably, in contrast to other WGM resonators, BMRs can be used as sensing devices where the medium under investigation is situated not only outside the resonator (for solid BMRs) but also inside and simultaneously outside and inside it (for hollow BMRs). The principle of sensing is based on the measurement of the resonance spectral shifts caused by the presence of the investigated medium. Sensing with WGM microresonators other than BMRs (e.g., spherical and toroidal resonators) can be similarly performed with BMRs. A review of BMR sensors was published recently [92]. Therefore, here we outline several studies on BMR sensing with hollow BMRs only, i.e., those related to microfluidic sensing.

Application of hollow BMRs to microfluidic sensing is similar to that of uniform microcapillaries [115]. Advantageously, hollow BMRs have much richer spectrum than uniform microcapillaries and in, several cases, can perform more comprehensive measurements. Characterization of fluids dwelling inside or propagating along the BMR axis is achieved by measurement and analysis of variation of the BMR resonant spectra. For example, the authors of Ref. [90] fabricated a hollow BMR illustrated in Fig. 21(a) and used it to probe the concentration of NIR-active dye (SDA2072) in methanol solution.  External and internal sensing with hollow BMR is illustrated in Fig. 21(b). The penetration of WGMs with small radial quantum numbers into the external environment is sufficient for its sensing. However, the internal penetration of these modes into liquid is negligible. For larger radial quantum numbers, the resonance spectrum of these modes is sensitive to variation of refractive index of liquids situated inside the BMR.  In Ref. [90], BMRs were fabricated from silica capillaries by HF etching followed by the softening and pressurizing procedure. BMRs with wall thicknesses between 5 and 10 μm were fabricated.  The experimentally measured spectra corresponding to different concentrations of dye are shown in Fig. 21(c) and demonstrate remarkably good sensitivity. It is seen that the major effect of the dye results in changing the intensity of the resonant dips, i.e., is related to losses.

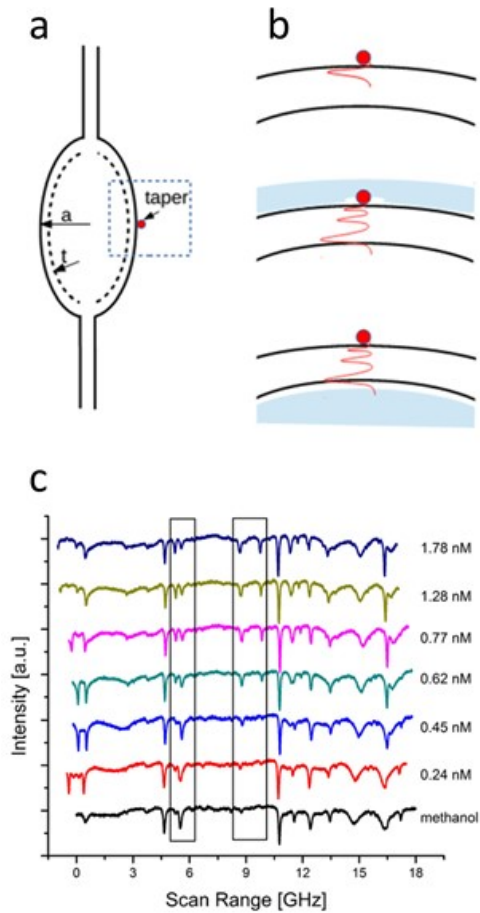

Fig. 21. (a) – Illustration of a hollow BMR. (b) – Comparison of external and internal evanescent sensing with WGMs depending on their radial quantum number. (c) – Resonant BMR spectra corresponding to different concentrations of dye (SDA2072) (Reproduced with permission from Ref. [90]).

One of the exciting applications of hollow BMRs for sensing explores their optomechanical properties [80-83]. As an example, Fig. 22(a) shows a hollow electro-opto-mechanical BMR demonstrated in Refs. [80, 81] enabling real-time detection of mechanical properties of microparticles flowing in liquid inside the BMR. The outer and inner diameters of BMR used in Ref. [81] were 70 μm and 50 μm. The microfluidic solution was composed of silica microparticles with approximately 3.62 μm diameter mixed in water. The 24.26 MHz frequency mechanical mode of this BMR was excited with an electrostatic drive (Fig. 22(a), right). Fig. 22(b) shows the frequency shifts of this vibrational mode as a function of time measured during transition of the microparticle solution inside the BMR. It is seen that the shifts associated with individual particle transits can be clearly resolved above the background noise fluctuations. Fig. 22(c) shows samplings of individual particle transits at flow rates of 50 μl/min. It has been shown that the particle speed, density and compressibility can be determined from these measurements.

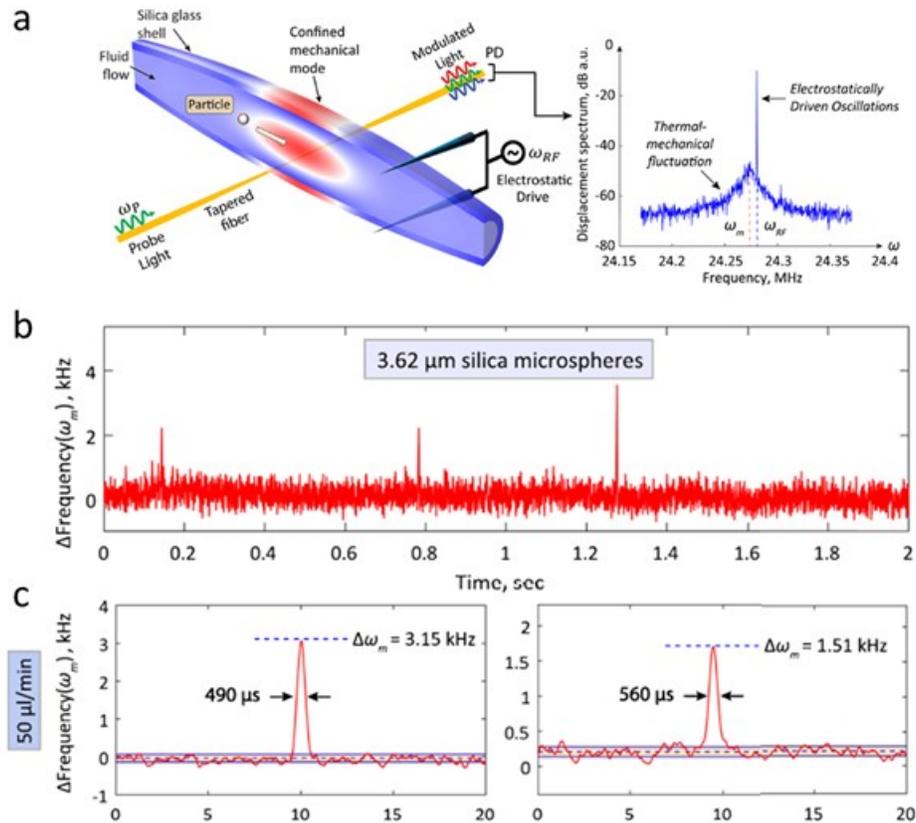

Fig. 22. (a) – Illustration of a confined mechanical mode in a hollow BMR (right) and the spectrum of its mechanical oscillations near the electrostatically driving frequency (left). (b) – The frequency variation of the output monitored in real-time using an oscilloscope. (c) – Magnified outputs

at the time of particle transition. The flow rate is 50 μl/min. (Reproduced with permission from Ref. [81]).

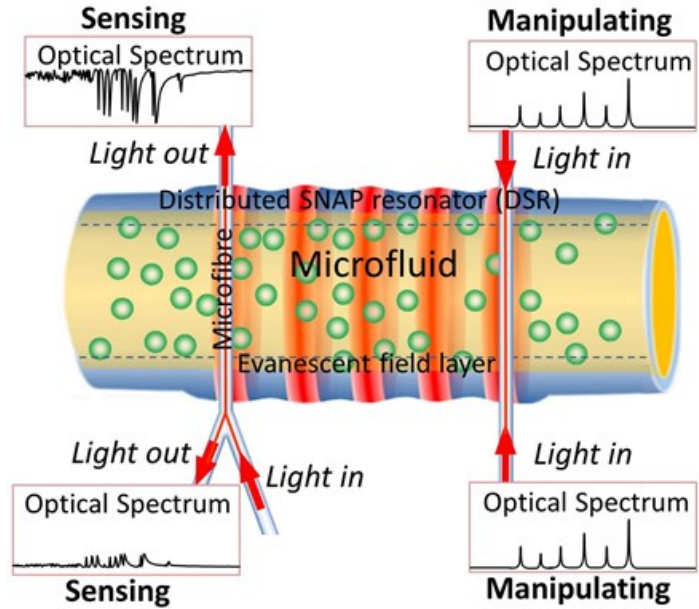

Fig. 23. Microfluidic sensor and manipulator. The spectrum of WGMs in the SNAP BMR introduced at the capillary surface can be used to restore the behavior of microfluidic components inside the capillary in time and space. While the left hand side microfiber is used to excite WGMs which are used for microfluidic sensing, the right hand side microfiber is used to excite WGMs which are used for microfluidic manipulation.

In contrast to WGM uniform fiber and uniform capillary sensors which can detect properties of adjacent medium localized along the BMR axis [115], the BMR sensing devices and SNAP microresonators elongated along the axis of an optical fiber enable sensing of medium distributed along the fiber axis [97]. Furthermore, the same microresonators can perform sensing of physical and chemical characteristics of liquids in microcapillaries and extract their spatial and temporal dependences and, simultaneously, manipulate the microfluidic processes with evanescent fields of WGMs localized in the introduced SNAP microresonators. The device enabling simultaneous sensing and manipulation of microfluidic processes is illustrated in Fig. 23. In this figure, the left hand side microfiber coupled to the SNAP resonator detects the positions of microparticles propagating along the capillary. It was suggested [97] that their positions can be found from the observed microresonator spectrum by solving the inverse problem. The second microfiber was proposed to be used to excite the resonant evanescent field which can control the microfluidic processes and, in particular, manipulate microparticles. The development of the complete theory of this device as well as its realization is believed to be feasible in the nearest future.

## 5. Summary

The elongated geometry of BMRs differentiates them from other WGM resonators having, e.g., toroidal or spherical shapes. Due to this geometry, BMRs possess several important properties which are not possible or difficult to achieve using other microresonators. For example, it was shown that a BMR with radius variation $\rho_w(z) = \rho_0 |\cos(\Delta kz)|$ (Eq. (2)) has frequency eigenvalues which are equally spaced (dispersionless) along the axial quantum number [11]. This property of BMRs is important for fabrication of miniature broadband and small repetition rate frequency comb generators proposed in [33]. BMRs with semi-parabolic nanoscale effective radius variation were demonstrated as record low loss, dispersionless, and small delay lines [24]. These microresonators can be fabricated with subangstrom-precise SNAP technology and potentially can be used as a basis for fabrication of miniature optical buffers [31].

Numerous fabrication methods of BMRs have been developed. They range from direct melting and pulling [18] and annealing [21] of the optical fiber to depositing curable polymer droplets [64] at the fiber surface, and rolling semiconductor films [38]. Different approaches for fabrication of hollow BMRs were developed as well, e.g., those based on pressurizing microcapillaries softened by a $CO_2$ laser beam [42].

Applications of BMRs include fabrication of microlasers [63, 64], nonlinear and optomechanical devices [69, 72, 82], quantum processors [87], ultraprecise microscopic sensors [90] and, in particular, microfluidic sensors [81]. Most of these applications employ the advantages of elongated geometry of BMRs. For example, this geometry made possible the demonstration of spatial beam engineering in fabrication of BMR microlasers [64] which can be further advanced employing SNAP BMRs [19]. As another example, of special interest are microfluidic BMR sensors [41] which enable monitoring the physical properties of liquids propagating along microchannels [81]. Potentially these sensors can be used for characterization as well as manipulation of microfluidic components [97, 27, 28]. The future development and applications of BMRs may boost based on the SNAP technology which enables fabrication of BMRs with complex shape and unprecedented subangstrom precision.